\documentclass[useAMS,usenatbib]{mn2e} 
\usepackage{txfonts,graphicx,natbib} 
\usepackage{longtable} 
\bibpunct{(}{)}{;}{a}{}{,} 
\newcommand{\myemail}{a.pastorello@qub.ac.uk} 
 
\title[Recurrent Outbursts in an Unusual LBV in NGC 3432]{Multiple Major Outbursts from a Restless Luminous Blue Variable in NGC 3432} 
\author[A. Pastorello et al.]{A. Pastorello$^{1}$\thanks{E-mail: \myemail}, M. T. Botticella$^{1}$, C. Trundle$^{1}$, S. Taubenberger$^{2}$,
S. Mattila$^{3,4}$, \and  E. Kankare$^{5,3}$, N. Elias-Rosa$^{6}$, S. Benetti$^{7}$,
G. Duszanowicz$^{8}$,  L. Hermansson$^{9}$, \and
J. E. Beckman$^{10,11}$, F. Bufano$^{7}$,  
M. Fraser$^{1}$, A. Harutyunyan$^{12}$, 
H. Navasardyan$^{7}$, \and  S. J. Smartt$^{1}$,  S. D. van Dyk$^{6}$, J. S. Vink$^{13}$, R. M. Wagner$^{14}$ \\ 
$^{1}$Astrophysics Research Centre, School of Mathematics and Physics, 
Queen's University Belfast, Belfast BT7 1NN, United Kingdom\\ 
$^{2}$ Max-Planck-Institut f\"{u}r Astrophysik, Karl-Schwarzschild-Str. 1, D-85741 
 Garching bei M\"{u}nchen, Germany\\ 
$^{3}$ Tuorla Observatory, Department of Physics \& Astronomy, University of Turku, V\"ais\"al\"antie 20, FI-21500 Piikki\"o, Finland\\ 
$^{4}$ Stockholm Observatory, Stockholm University, AlbaNova University Center, SE-10691, Stockholm, Sweden\\
$^{5}$ Nordic Optical Telescope, Apartado 474, E-38700 Santa Cruz de La Palma, Spain\\
$^{6}$ Spitzer Science Center, California Institute of Technology, 1200 E. California Blvd., Pasadena, CA 91125, USA\\ 
$^{7}$ INAF Osservatorio Astronomico di Padova, Vicolo dell'Osservatorio 5, I-35122 Padova, Italy \\ 
$^{8}$ Moonbase Observatory, Otto Bondes vag 43, SE-18462 Akersberga, Sweden\\ 
$^{9}$ Sandvretens Observatory, Linn\'egatan 5A, SE-75332 Uppsala, Sweden \\ 
$^{10}$ Instituto de Astrof\'{i}sica de Canarias C/ V\'{i}a L\'actea s/n, E-38205, La Laguna, S/C de Tenerife, Spain\\ 
$^{11}$ Consejo Superior de Investigaciones Cient\'{i}cas of Spain, C/ Serrano 117, E-28006, Madrid, Spain\\ 
$^{12}$ Fundaci\'on Galileo Galilei - INAF, Telescopio Nazionale Galileo, Rambla Jos\'e Ana Fern\'andez P\'erez 7, E-38712 Bre\~na Baja, Tenerife, Spain\\ 
$^{13}$ Armagh Observatory, College Hill, Armagh BT61 9DG, United Kingdom\\
$^{14}$ Large Binocular Telescope Observatory, 933 North Cherry Avenue, Tucson, AZ 85721, USA\\ 
} 

\begin{document} 
 
\date{Accepted .... Received .....; in original form .....} 
 
 
\maketitle 
 
\label{firstpage} 
 
\begin{abstract} 
We present new photometric and spectroscopic observations of an unusual luminous blue variable (LBV)
in NGC 3432, covering three major outbursts in October 2008, April 2009 and November 2009. Previously, this star experienced 
an outburst also in 2000 (known as SN 2000ch). During outbursts the star reached an absolute magnitude between -12.1 and -12.8.
Its spectrum showed H, He I and Fe II lines with P-Cygni profiles during and soon after the eruptive phases,
while only intermediate-width lines in pure emission (including He II $\lambda$ 4686) were visible during quiescence.
The fast-evolving light curve soon after the outbursts, the quasi-modulated light curve, 
the peak magnitude and the  
overall spectral properties are consistent with multiple episodes of variability of an 
extremely active LBV. However, the widths of the spectral lines indicate unusually high  
wind velocities (1500-2800 km s$^{-1}$), similar to those observed in Wolf-Rayet stars. 
Although modulated light curves are typical of LBVs  
during the S-Dor variability phase, the luminous maxima
and the high frequency of outbursts are unexpected in S-Dor variables.
Such extreme variability may be associated with
repeated ejection episodes during a giant eruption of an LBV.
Alternatively, it may be indicative of a high level of instability shortly preceding the core-collapse 
or due to interaction with a massive, binary companion. In this context, the variable in NGC 3432 shares some similarities 
with the famous stellar system HD 5980 in the Small Magellanic Cloud, 
which includes an erupting LBV and an early Wolf-Rayet star.

\end{abstract} 
 
\begin{keywords} 
stars: variables: other - supernovae: general - supernovae: individual: 2000ch - galaxies: individual: NGC 3432. 
\end{keywords} 
 
\section{Introduction} 
 
Luminous Blue Variable (LBV) stars are  debated subjects of modern astrophysics,  
because they signify a key stage in the fate of  
the most massive stars. Such stars may experience an unstable post-main-sequence LBV stage
before becoming  Wolf-Rayet (WR) stars. LBVs lose   most of the hydrogen (H) envelope
rapidly \citep[typically 10$^{-5}$ M$_\odot$ yr$^{-1}$, up to 10$^{-4}$ M$_\odot$ yr$^{-1}$, see
e.g.][]{hum94}, forming massive circumstellar nebulae like the  
spectacular {\sl Homunculus} surrounding $\eta$-Car.
Later  massive stars eventually  explode as type Ib/c supernovae (SNe) \citep[see the case of SN 2001am,][]{sci08}. However,  
there is increasing evidence that even LBVs can {\it directly} explode  
producing core-collapse SNe \citep{kot06,gal07,gal08,smi07,smi08,tru08,agn08}. 
 
 An attempt to characterize LBVs was done by \citet{hum94}. 
However, LBVs do not form a homogeneous group of stars,  
since they span quite a large range of magnitudes and variability types. Some of them show an S-Dor variability type.  
During this phase, the star expands and contracts with some regularity over timescales of years, varying  
its apparent temperature (and, as a consequence, spectral type). Its bolometric luminosity is generally
not thought to change significantly \citep[but see][]{cla09,gro09} and the upper limit to the 
luminosity/mass ratio for static stellar atmospheres (the so-called ``Eddington limit'') is not violated.
With S-Dor, other classical examples of this variability type are AG Car in the Galaxy and R 127 in the Large Magellanic Cloud (LMC).
As mentioned by \citet{vge01}, the S-Dor variability has typical periods of few years ({\it short S-Dor phase}) or decades ({\it long S-Dor 
phase}). Other LBVs occasionally experience giant eruptions, during which they increase their luminosity and can  
reach $M \approx$ -14 (4-6 mag brighter than their typical quiescence magnitudes), temporarily exceeding the Eddington Limit.
The most popular giant eruption of an LBV in our Galaxy is that of $\eta$-Car during the 19th century.
  
Spectra of LBVs' giant eruptions are characterized by incipient narrow (with full-width-at-half-maximum
velocity, $v_{FWHM}$, lower than 1000 km s$^{-1}$) H lines in emission, resembling  
those of type IIn SNe. For this reason and their relatively high peak luminosities,
they are occasionally misclassified as real SN explosions.
However, since the stars survive the eruptions, they are  labelled as {\it SN impostors} 
\citep{van00}. Well-studied cases are SN 1954J \citep[also known as V12 in NGC 2403,][]{tam68,smi01,van05},
SN 1997bs \citep{van00} and  SN 2002kg \citep{wei05,mau06,mau08}\footnote{Many other transients have been proposed to be 
SN impostors, including SN 1999bw \citep{fil99,li02}, SN 2001ac \citep{mat01}, 
SN 2003gm \citep{mau06}, SN 2006fp \citep{blo06}, SN 2007sv \citep{har07}, SN 2009ip \citep{smi09b,fol10}.
In addition, the controversial SN 1961V, whose nature (SN explosion or LBV outburst) is still debated  
\citep{utr84,fes85,bra85,cow88,goo89,fil95,sto01,van02,chu04} deserves to be mentioned.}.
The first registered outburst of the 
 variable star in NGC 3432 discussed in this paper was designated as SN 2000ch \citep[][]{pap00,wag04}.
Hereafter, this transient will be dubbed as NGC 3432 2000-OT.  
Contrary to genuine SNe IIn, SN impostors are less luminous and their spectra do not show clear evidence of broad lines ($>$ 3000-4000  
km s$^{-1}$) produced by SN ejecta. SN impostors will hereafter be referred to
without the misleading ``SN'' label.  
  
In this paper we will analyse the October 2008, April 2009 and November 2009 outbursts of the same variable in NGC 3432 that was responsible  
for NGC 3432 2000-OT \citep[see the wide discussion in][]{wag04}.  
We attempt  to reconstruct the variability history of this star and derive information on its nature. 
The paper is organized  
as follows: in Sect. 2 we introduce the variable in NGC 3432 and describe its environment. In Sect. 3 and Sect. 4  
photometric and spectroscopic evolutions are presented, while in Sect. 5 we analyse the evolution of the  
spectral energy distribution (SED) of the variable. 
A discussion on the nature of the star follows in Sect. 6. 
  
\section[]{The luminous variable in NGC 3432 and its environment} 
   
\begin{figure} 
\includegraphics[width=84mm,angle=90]{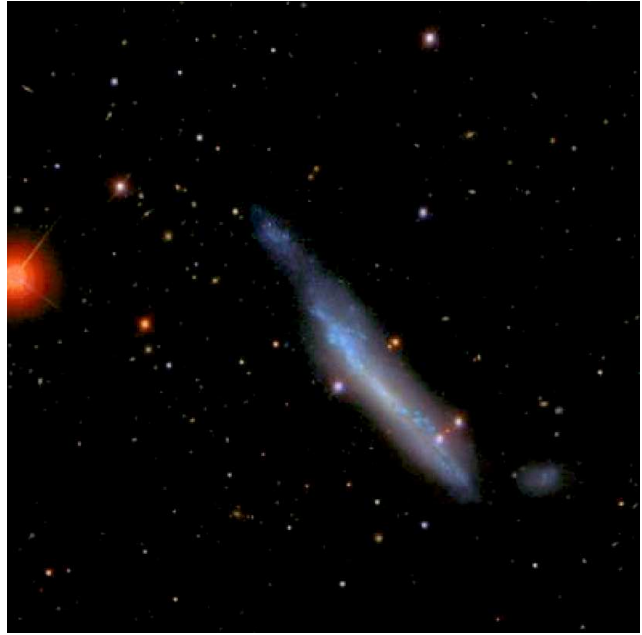} 
  \caption{SDSS colour image ({\it http://cas.sdss.org/astrodr6/en/tools/chart/}) of the interacting galaxy system Arp 206.
The main galaxy is NGC 3432, while the dwarf companion UGC  5983 is located south-west of it.  
The field of view is about 11$\arcmin \times $11$\arcmin$. North is up, East is to the left.} \label{fig0} 
\end{figure} 
 
NGC 3432 is a well-studied irregular galaxy \citep[see e.g.][]{ho97} interacting with a low-surface brightness dwarf companion (UGC 5983), 
which is located South-West of the main galaxy (Figure \ref{fig0}). 
This interacting system is included in Arp's catalogue \citep{arp66} with the label Arp 206. 
\citet{eng97} noted the presence of an extended arm of the galaxy in the North-East direction, which ends with 
a luminous clump in which the variable star lies. The disturbed morphology of NGC 3432 and the presence of an extended North-East region  
detached from the main body of the galaxy likely result from the tidal interaction of NGC 3432 with UGC 5983 
\citep[see also][]{swa02,vdk07}. 

 \citet{lee09} determined for NGC 3432 a star formation rate (SFR) of 0.49 M$_\odot$ yr$^{-1}$ from UV data, and of 0.32 M$_\odot$ yr$^{-1}$ from H$\alpha$ luminosity.
Interaction between galaxies is well known to trigger star formation, and  
the relatively low SFR observed in NGC 3432 may appear as an inconsistency. However, a starburst episode may occur with a significant delay  
(10$^8$-10$^9$ yrs) with respect to the major tidal distortions \citep{mih94}. According to \citet{eng97}
Arp 206 is consistent with this scenario, and would currently be in a pre-starburst phase.  
  
Measuring the intensity of the [O II] $\lambda\lambda$ 7320,7330 doublet and the intensity ratio of  
strong [O II] $\lambda$ 3727 and [O III] $\lambda\lambda$ 4959,5007 lines in H II regions, \citet{pyl07} estimated an average O  
abundance of 12 log (O/H) = 8.3 $\pm$ 0.1 dex for NGC 3432 which is comparable to the metallicity of the LMC 
\cite[$\sim$8.35 dex,][]{hun07}. However, the metallicity in the region of the variable star might 
be significantly different from this average value. 
  
The distance of the galaxy system is computed through the recessional velocity corrected for  Local Group infall into 
Virgo\footnote{The Local Group infall velocity here adopted is $v_{LG-infall}$ = 208 $\pm$ 9  km s$^{-1}$ \protect\citep{ter02}}. The Hyperleda 
database\footnote{http://leda.univ-lyon1.fr/} gives $v_{Vir}$ = 779 km s$^{-1}$, 
providing a distance of 10.8 $\pm$ 1.2 Mpc 
\citep[distance modulus $\mu$ = 30.17 $\pm$ 0.56 mag, adopting $H_0$ = 72 $\pm$ 8 km s$^{-1}$ Mpc$^{-1}$, ][]{fre01}. 

\citet{sch98} quote a Galactic reddening of
$E(B-V)$= 0.013 in the direction of NGC 3432, while the peripheral position of the objects in the host galaxy suggests 
a modest internal reddening. This is also indicated from the lack of narrow interstellar absorption lines in the spectra of the variable star (see Sect. \ref{sp}).
Therefore, hereafter we will adopt a total reddening of $E(B-V)$= 0.013 mag, in agreement with \citet{wag04}. 
 
The discovery of a variable star in NGC 3432 on 2000 May 3 UT was reported by \citet{pap00}  
at the position $\alpha = 10^{h}52^{m}41\fs40$ and $\delta = +36\degr40\arcmin08\farcs5$ (equinox J2000.0), which is 123$\arcsec$ East and 180$\arcsec$ 
North of the nucleus of the host galaxy. The transient  
had an unfiltered peak magnitude of $m$ = 17.4. A KAIT image was obtained 4 days prior to this on 2000 April 29 showing nothing brighter than $m$ =  19 at the position  
of the variable \citep{wag04b}. The low luminosity and the spectra dominated by 
narrow Balmer lines were consistent with a SN impostor \citep[e.g.][]{van00}.  
Wagner et al. presented optical plus near-infrared light curves of  NGC 3432 2000-OT covering a period of  
$\sim$3 months, and noted a somewhat erratic variability over very short timescales (of the order of weeks). 
Additional sparse archive observations led these authors to conclude that no major outburst was previously observed,  
and the likely quiescent magnitude of the LBV was around $R$ = 19.4 $\pm$ 0.4 mag. They also registered a deep magnitude  
minimum ($R \sim$ 20.8) soon after the luminosity peak, and interpreted it as the result of dust formation, occultation or  
eclipse phase. We will show in Sect. \ref{ph} that the star has been detected (occasionally) at a magnitude even fainter than $R \approx$ 21 
in subsequent follow-up observations.  
 
\begin{figure} 
\includegraphics[width=82mm]{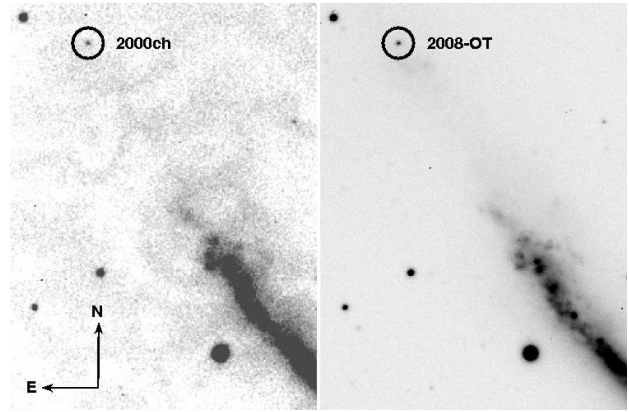} 
  \caption{Comparison between an unfiltered KAIT image of NGC 3432 2000-OT (2000ch) obtained on May 4, 2000 and 
  a R-band image of the 2008 transient obtained at TNG on 25 October, 2008. The position of the two
  transients is coincident, within the errors. This allows us to conclude that the two outbursts were produced by the same star.} \label{fig1} 
\end{figure} 
 
\begin{figure} 
\includegraphics[width=82mm]{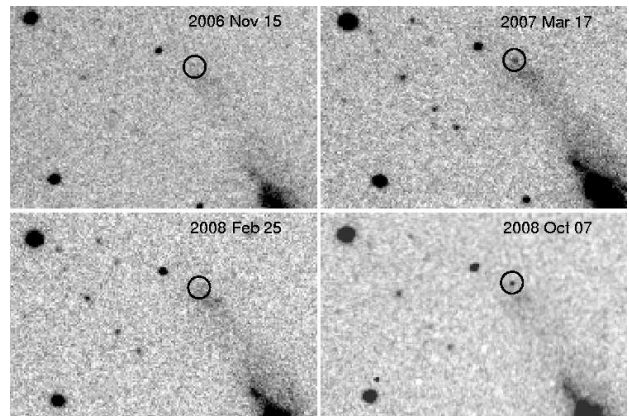} 
  \caption{Unfiltered images of NGC 3432 obtained with a 32-cm reflector in Akersberga (Sweden). The 4 images show the evolution of the variable star, whose position is marked with a circle, 
on 2006 November 15 (top-left), 2007 March 17 (top-right), 2008 February 25 (bottom-left) and 2008 October 7 (bottom-right). 
The transient is clearly visible in the March 2007 and October 2008 images, while in the other two images it is below the detection 
limit. North is up, East is to the left.} \label{fig2} 
 \end{figure} 
 
Another transient was announced on 2008 October 7.12 UT by \cite{dus08} at a position $\alpha = 10^{h}52^{m}41\fs33$, 
$\delta = +36\degr40\arcmin08\farcs9$ (equinox J2000.0),  
very close (but apparently not coincident) to that of the 2000 outburst (Figure \ref{fig1}).  
The new transient (labelled as NGC 3432 2008-OT in this paper) had an apparent peak magnitude ($R$ = 17.5)  
similar to that of 2000-OT. Its faint absolute magnitude  ($M_R$ $\approx$ -12.7) 
was again consistent with that of a SN impostor. 

In order to prove that the position of 2008-OT was coincident with that of 2000-OT, 
relative astrometric calibration was performed using as template 
a good-quality image of the 2008 transient and as target image the discovery frame
of the 2000 transient obtained with the 0.76-m KAIT telescope (Filippenko et al. 2001).
In order to align the two images, we identified 15 sources in common to both images and measured their centroid 
positions. This allowed us to derive a geometric transformation for the two images (implying a shift, a scaling 
to a common pixel scale and a rotation) using the IRAF package GEOMAP. To employ a general non-linear transformation, 
we selected a 2nd-order polynomial model that computes the geometric alignment function. The images were finally
registered to the target image using GEOTRAN. The accuracy of the alignment is 0.145 px in the X-axis and
0.154 px in Y. Accounting for the pixel scale of the target image, 
the rms errors of the transformation in the two axes are 116 mas and 123 mas.
We measured the position of the transients in the two images with  aperture photometry and found 
in the transformed images a discrepancy in the position of the two transients of $\Delta x$ = 0.039 px (31 mas)
and $\Delta y$ = 0.013 px (10 mas), which are well below the uncertainty of the transformation.
We are therefore confident that the two transients were two major outbursts of the same stellar source (Figure \ref{fig1}). 

\begin{figure*} 
\includegraphics[width=154mm,angle=0]{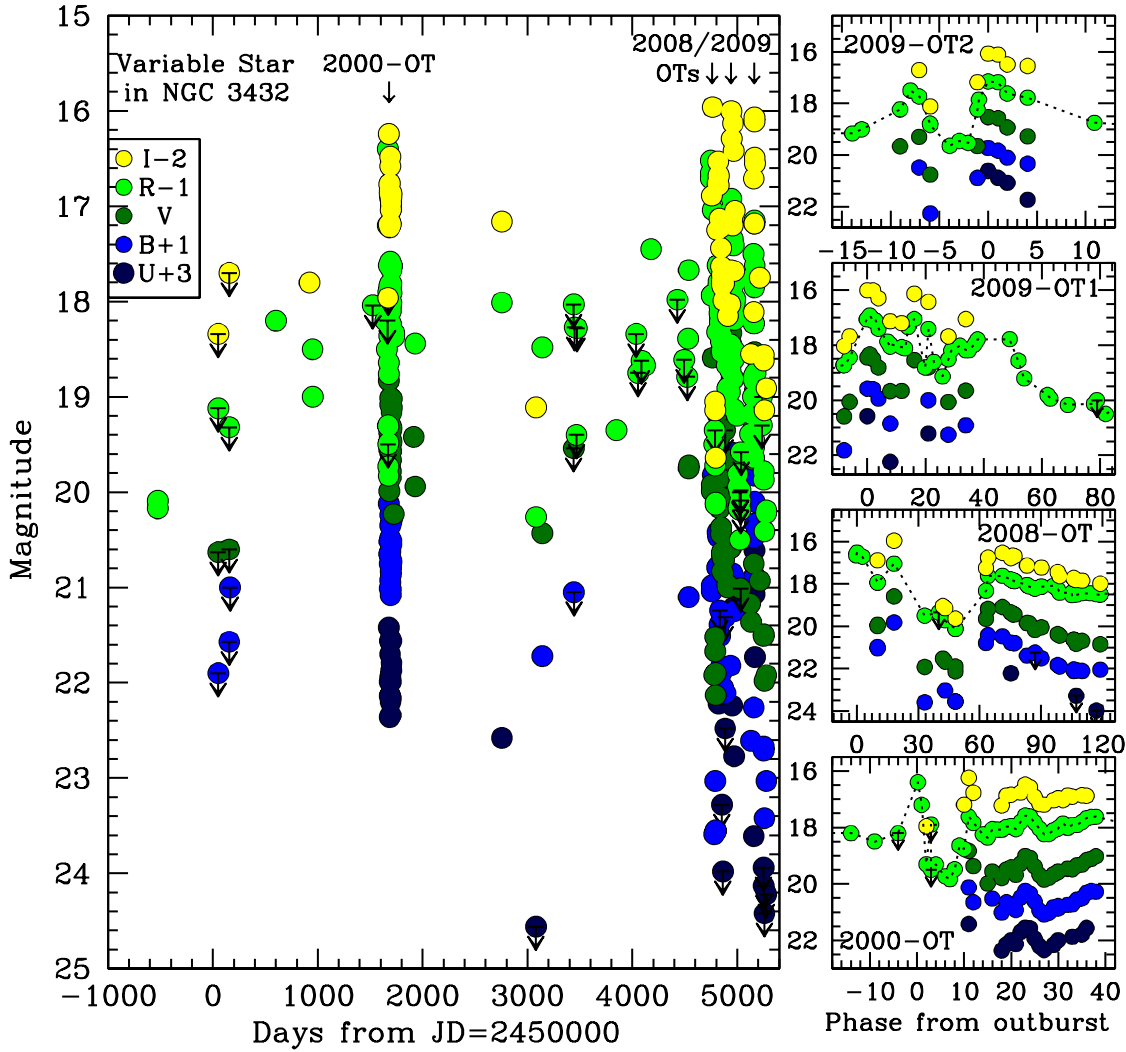} 
  \caption{Main panel: 15-yrs multi-band optical light curve of the variable in NGC 3432. Blow-ups with the detailed light curves of  
  2000-OT \protect\citep[the data are from][]{wag04}, 2008-OT, 2009-OT1 and 2009-OT2 are shown  
  in the panels on the right. Sloan u', r' and i' magnitudes were scaled to 
  Johnson-Bessell U, R and I magnitudes applying a correction of $\Delta U$ = -0.81,  $\Delta R$ = -0.18 and $\Delta I$ = -0.41 mag, computed  
  comparing the photometry of the local sequence stars in the Sloan and Johnson-Bessell photometric systems. To guide the eye,  
  R-band data points in the inserts are connected by a dotted line.} \label{fig3} 
\end{figure*} 

More precise coordinates than those previously reported for the NGC 3432 variable, as determined from Sloan Digital Sky Survey (SDSS) images, are  
$\alpha = 10^{h}52^{m}41\fs256$ and $\delta = +36\degr40\arcmin08\farcs$95 (equinox J2000.0). 
Unfiltered amateur images showing the region of the variable star at 4 epochs during the period 2006-2008 are shown in Figure \ref{fig2}. 
Finally, two further outbursts of the variable in NGC 3432 were registered in April 2009 (hereafter labelled as 2009-OT1) and November 2009 (2009-OT2), 
during routine monitoring observations of the star. 
    
\section[]{Photometry} \label{ph} 
 
After the discovery by \citet{dus08}, we started a systematic monitoring of the variable in the optical bands.  
At the same time, we collected a number of pre-discovery images of NGC 3432 from amateur astronomers and professional  
telescopes available through science data archives. The calibration of the optical photometry obtained with standard Johnson-Bessell 
filters has been performed using the local standard star magnitudes presented in \citet{wag04} for the B, V and R bands, while the U and I
band data have been calibrated independently. The calibration of the Sloan filter photometry was performed using 
the SDSS photometry catalogue\footnote{Sloan Digital Sky Survey web site: {\it 
  http://www.sdss.org/}. The SDSS filter definition can be found in \citet{fuk96}, while the photometric system is defined in \citet{smt02}}.  
 
\begin{figure*} 
\includegraphics[width=100mm,angle=270]{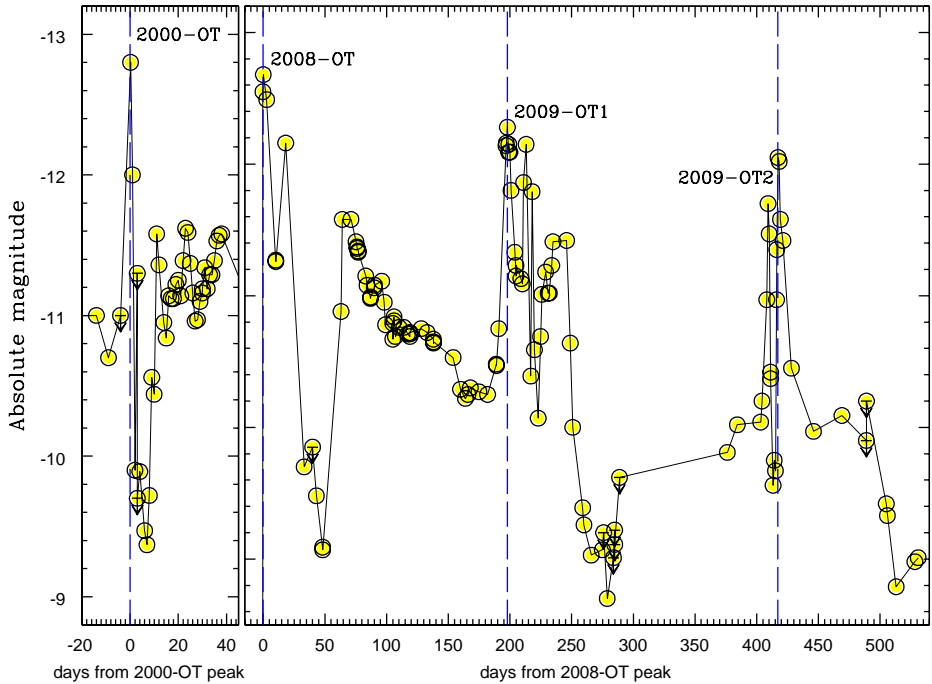} 
  \caption{Comparison of the R-band light curves of the variable star in NGC 3432 covering
  the 4 major outbursts:  2000-OT \protect\citep{wag04}; 2008-OT, 2009-OT1, 2009-OT2 (this paper). 
  SDSS r-band magnitudes were scaled to Johnson-Bessell ones by applying a  
 zero-point shift of $\Delta R$ = -0.177 mag (see caption in Figure \ref{fig3}). The epochs of the main luminosity peaks
 for the 4 outbursts are marked with vertical blue dashed lines.} \label{fig4} 
\end{figure*} 

Johnson-Bessell and Sloan magnitudes (or magnitude limits) of the variable in NGC 3432 are reported 
in Appendix (Tables \ref{tab1} and \ref{tab2}). 
The light curves, including the photometry of \citet{wag04} and  
spanning a period of almost 16 years, are shown in Figure \ref{fig3}. The right-hand panels show in detail the evolution of the 
variable during the 4  
major outbursts: from top to bottom 2009-OT2,  2009-OT1, 2008-OT and 2000-OT. As already mentioned, the outbursts  
showed comparable peak magnitudes ($R \approx$ 17.5-18.5 mag) and a number of subsequent luminosity fluctuations,  
although the post-peak evolution is somewhat different in the 4 cases. This is clearly visible when the R band absolute light curves  
of the 4 outbursts are plotted together (Figure \ref{fig4}). 
While 2000-OT showed a fast post-peak decline, with a deep minimum reached after a few days, 
 2008-OT and 2009-OT1 reached the minimum 2-3 months after maximum. 
We can summarize the photometric evolution of the 4 outbursts as follows: 

\begin{itemize} 
 \item {\bf 2000-OT.} During this outburst, the star reached a peak magnitude of $R$ = 17.4 ($M_R \approx$ -12.8) on May 3rd, 2000. The maximum was followed by a steep decline of 3.5 mag 
 in about a week (but an observation obtained 2 days after maximum already showed the transient being 2.9 mag fainter than at peak). 
 A sequence of secondary peaks was also observed (at magnitude around 18.6), with a temporal distance of about 12-15 days between each other  
 \citep[][and Figure \ref{fig3}, bottom-right panel]{wag04}.  
 \item {\bf 2008-OT.} During this eruptive episode, the variable arrived at a maximum magnitude of $R$ = 17.5 ($M_R \approx$ -12.7, 
 October 7th, 2008), and experienced  
 a rebrightening ($R \approx$ 18) at $\sim$18 days after the main maximum. The rebrightening   
 was followed by a large luminosity drop (more than 3 mag). Although the light curve is poorly sampled in this 
 period, the star probably remained fainter than $R \sim$ 20.5 ($V \sim$ 21.5) for about 1 month.  
 Then its luminosity increased again, reaching a peak of $R \sim$ 18.6 at about 70 days after  
 the main outburst. Finally the luminosity slowly declined and leveled off at $R \approx$ 19.5 (see Figure \ref{fig3}, right, second panel from
 bottom).  
 \item {\bf 2009-OT1.} Surprisingly, about 200 days after the previous outburst the variable brightened again. 
 In this episode the peak magnitude (on April 22nd, 2009) was around $R \sim$ 17.9 ($M_R \sim$ -12.3). The primary peak was  followed by a number of magnitude oscillations that 
 occurred on shorter time-scales than those observed in 2008-OT. They were
followed by a dramatic drop of $\Delta R$ $\geq$ 3.3 mag from the peak to the minimum (Figure \ref{fig4}). 
 \item {\bf 2009-OT2.} A further rebrightening of the variable in NGC 3432 was finally registered in November 2009, again {\bf $\sim$}200-210 days after the previous
 episode. This might be either a coincidence in the luminosity fluctuations or, alternatively, might suggest
 some periodicity in the major episodes of stellar variability. The consequences of this latter scenario will be discussed in Sect. \ref{prog}.
 Interestingly, the first maximum of 2009-OT2 reached  $M_R \sim$ -11.7, but was followed after 8 days by 
 a second peak that was even slightly brighter ($M_R \sim$ -12.1, see Figure \ref{fig4}). 
 After the second  peak, the luminosity of the star dropped down by more than 2 mag in $\sim$1 month.  
\end{itemize} 
 
\begin{figure} 
\includegraphics[width=87mm]{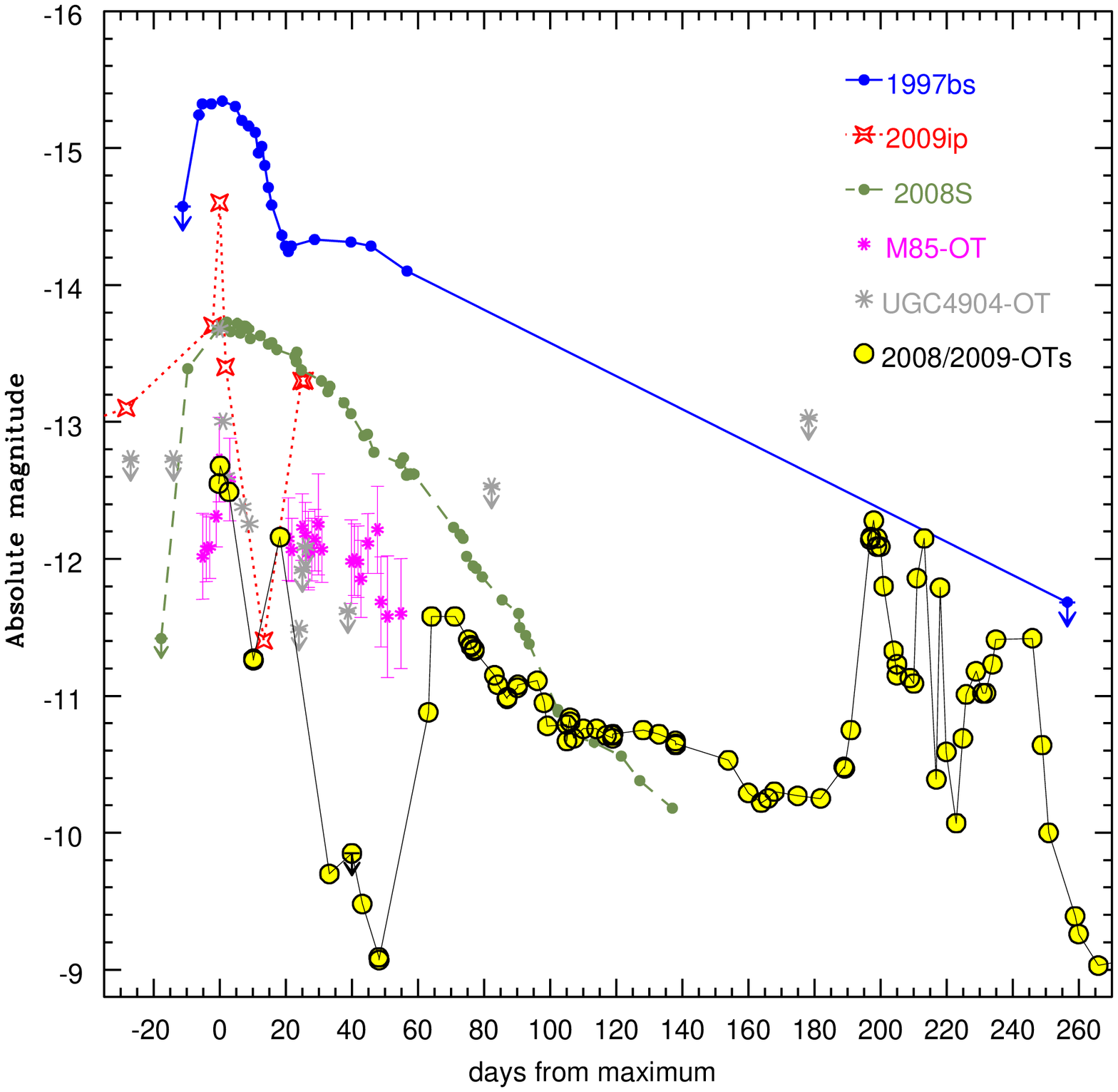} 
  \caption{Comparison of the R-band absolute light curve of  
  the variable star in NGC 3432 (period october 2008 - June 2009) with those of some well-studied transient events:  
 1997bs \protect\citep{van00}, 2009ip \protect\citep{smi09b}, the 2004 outburst preceding SN 2006jc \protect\citep[UGC 4904
 OT 2004-1,][]{pasto07a}, 2008S \protect\citep{bot08} 
 and M85 OT 2006-1 \protect\citep{kul07}. SDSS r-band magnitudes of the variable in NGC 3432 were rescaled to Johnson-Bessell magnitudes by applying a  
 zero-point shift of $\Delta R$ = -0.177 mag (see caption in Figure \ref{fig3}).} \label{fig5} 
\end{figure}

The comprehensive light curve (period 1994-2009, see Figure \ref{fig3}, main panel) shows a rather erratic  evolution. 
Luminosity peaks occur on unusually short timescales and the range of spanned magnitudes 
($\Delta m$ $\sim$ 4 mag, Figure \ref{fig4}) are atypical for an S-Dor-like event.  
Together with the main peaks we also note a few sparse detections at magnitudes 18.5-19.5, and some minima (e.g. at JDs 2449475 and 
 2453081). Such optical deficits can be possibly
due to post outburst dust formation episodes  \citep[as proposed by][]{wag04}. The implications of this unusual photometric evolution in constraining the nature of the variable star will be discussed in Sect. 5. 

In Figure \ref{fig5} we compare the R band absolute light curve of the variable in NGC 3432  during the period from October 2008 to June 2009 
with those of the SN impostors 1997bs \citep{van00} and 2009ip \citep{smi09b}, the 2004 luminous eruption in UGC 4904 that heralded SN 2006jc 
 \citep[hereafter UGC 4904 OT 2004-1,][]{pasto07a} and  
two additional under-luminous transients whose nature (core-collapse SN, LBV outburst, peculiar luminous nova) is  
still debated: M85 OT 2006-1 and 2008S \citep{kul07,rau07,pasto07b,ofek08,pri08a,tho08,wan08,bot08,smi09}.  
The peak magnitudes of the NGC 3432 variable (in the range -12.1 to -12.8) are much fainter than those of
1997bs ($M_R \approx$ -15.3) and 2009ip ($M_R \approx$ -14.6), marginally fainter than those of 2008S and UGC 4904 OT 2004-1 ($M_R \approx$ -13.7). 
However, they are comparable with the maximum magnitude of  M85 OT 2006-1 \citep{kul07}, although the two transients have different photometric evolution.  
M85 OT 2006-1, indeed, presents a slowly evolving light curve with a sort of plateau, while the variable in NGC 3432 has a fast evolution, 
characterized by steep post-peak luminosity declines and subsequent
rebrightenings similar to those observed in 2009ip \citep{smi09b} and other SN impostors \citep{mau06}.

\begin{figure*} 
\includegraphics[width=126mm,angle=270]{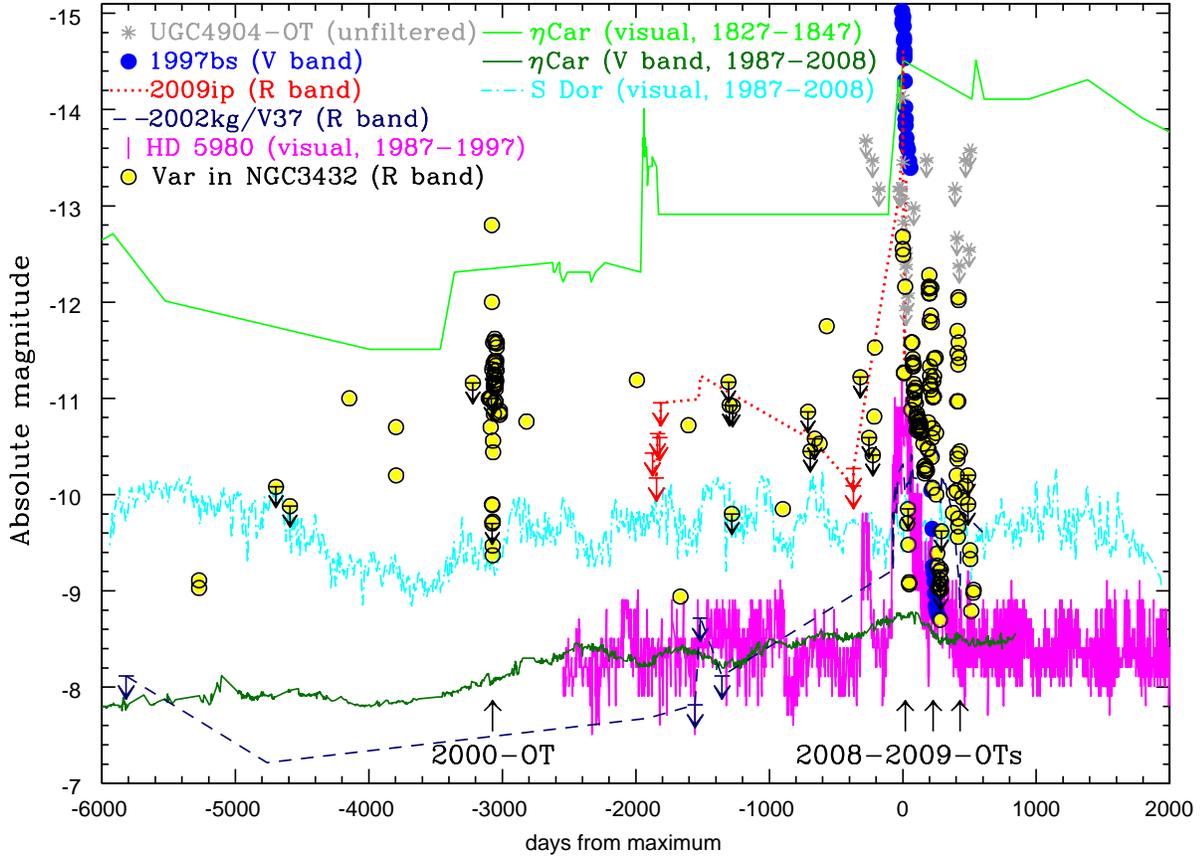} 
  \caption{Long-term photometric monitoring of a few famous LBV outbursts - SN impostors: 1997bs, 2000kg/V37, 2009ip, 2000/2008/2009-OTs in NGC 3432 and UGC 4904 OT 2004-1 
  (references in the text). The Sloan r-band magnitudes of the variable in NGC 3432 were rescaled to Johnson-Bessell magnitudes  
  adopting the prescriptions mentioned in the caption of Figure \ref{fig4}. A boxcar filter of size = 5 days was applied to the light curve of S-Dor (AAVSO database).
  The modulated variability of S-Dor, and the 
  light curve of the LBV/WR system HD 5980 in the SMC (including photometry of the eruption in 1993-1994, magenta crosses, AAVSO database) are also shown.  
  Visual magnitudes of $\eta$ Car during the Great Outburst of the 
  1840s  are visualized with green circles \protect\citep[see][and references therein]{fre04}, while 
  V band magnitudes of $\eta$ Car from 1987 to the present days  
(S. Otero, private communication) are shown with green-olive dots. The light curve of the Great Eruption of $\eta$ Car has been corrected 
for a total extinction of $A_V$ = 1.7 mag \protect\cite[e.g.][]{dav97}, while for the light curve in more recent decades we accounted for circumstellar 
dust, and therefore adopted a significantly higher extinction \protect\cite[$A_V$ = 6.1 $\pm$ 0.6 mag, according to][]{dav95}.} \label{fig6} 
\end{figure*}

In Figure \ref{fig6}  long-term light curves of the NGC 3432 variable and a few   
well monitored LBVs and SN imposters are shown: $\eta$ Car\footnote{Historical observations of  $\eta$ Car's Great Eruption in the 1840s were  
kindly provided by \protect\citet{fre04}, while the collection of the normalized V band photometric studies of the $\eta$ Car's variability  
after 1952 are from S. Otero (private communication). Extensive light curves of $\eta$ Car can be found at the following URL address: 
{\it http://varsao.com.ar/Curva$_{-}$Eta$_{-}$Carinae.htm}},  
S-Dor and HD 5980\footnote{Data from AAVSO International Database: {\it http://www.aavso.org}}, 
1997bs \citep{van00}, 2002kg \citep{wei05,mau06,mau08}, 2009ip \citep{smi09b}, UGC 4904 OT 2004-1 \citep{pasto07a}. 
Remarkably, even though some modulation in the light curve of the variable star in NGC 3432  
may suggest an S-Dor-type variability, its overall characteristics are 
reminiscent of those of major outbursts. Some similarity can be found, indeed, with the years-long outburst of $\eta$ Car during the 
19th century, and -more marginally- with the eruption in 1993-1994 of the stellar system HD 5980 \citep[see][and references therein]{koe04}. 
We will better address the latter similarity in forthcoming sections. 
 
\section[]{Spectroscopy} \label{sp} 
  
\begin{table*} 
 \centering 
 \begin{minipage}{165mm} 
  \caption{Basic information on new spectra of the luminous variable in NGC 3432.} \label{tab3} 
  \begin{tabular}{ccccccccc} 
  \hline 
Date & JD  &  Instrument & Grism/grating & Exposure time (s) & Range (\AA) & Resolution (\AA) \\ \hline 
2008Oct14 & 2454753.75 & WHT+ISIS & R300B,R158R & 2$\times$2400 & 3250-5410,5140-10150 & 5.4,6.3 \\ 
2008Oct17 & 2454756.71 & CAHA2.2m+CAFOS & b200 & 2727.3 & 3400-8770 & 10.5 \\ 
2008Oct17 & 2454756.74 & TNG+LRS & LRB & 2700 & 3700-7990 & 15 \\  
2008Dec21 & 2454821.75 & NOT+ALFOSC & gm4 & 3600 & 3300-9050 & 18 \\ 
2008Dec22 & 2454822.63 & TNG+LRS & LRR & 3600 & 5050-9800 & 13 \\ 
2009Jan22 & 2454853.64 & TNG+LRS & LRB & 3600 & 3560-7920 & 12 \\ 
2009Feb22 & 2454884.56 & NOT+ALFOSC & gm4 & 2$\times$2700 & 3400-8100 & 18 \\ 
2009Mar22 & 2454912.52 & TNG+LRS & LRB & 2$\times$2700 & 3370-7930 & 15 \\ 
2009Apr24 & 2454945.60 & TNG+LRS & LRB & 3600 & 3260-7940 & 11 \\ 
2009May12 & 2454964.49 & TNG+LRS & LRB & 2700 & 3280-7940 & 11 \\ 
2009May19 & 2454971.41 & TNG+LRS & LRB & 2700 & 3400-7930 & 11 \\ 
2009Nov27 & 2455162.67 & CAHA2.2m+CAFOS & b200 & 2700 & 3500-8730 & 12 \\
\hline 
\end{tabular} 
\end{minipage} 
\end{table*} 

\begin{figure*} 
\includegraphics[width=165mm]{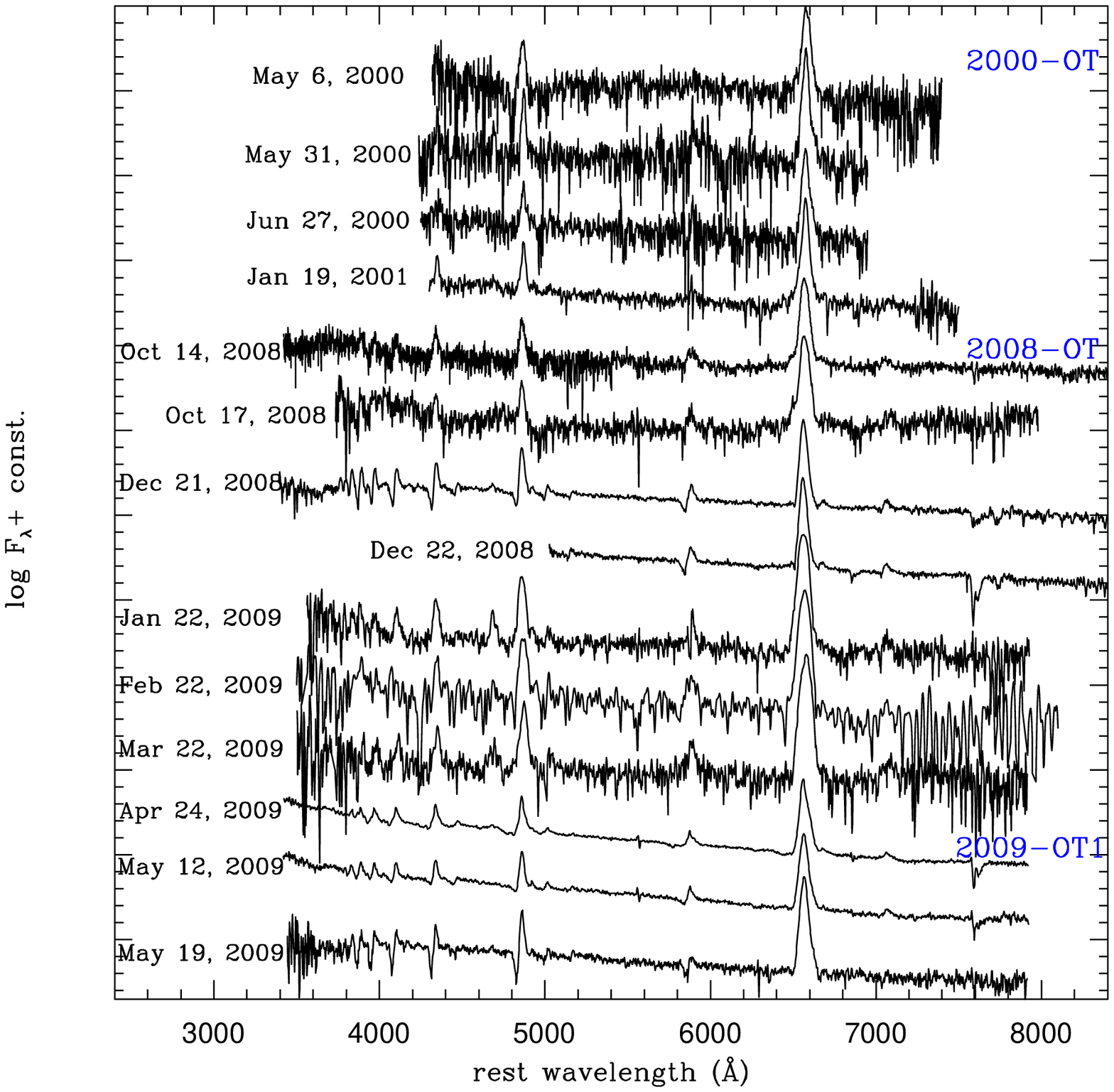} 
  \caption{Optical spectra of the variable in NGC 3432 from \protect\cite{wag04} (relative to the 2000 eruptive episode) and new spectra of the  
  2008/2009 outbursts. Only the best S/N spectra are shown. The second order contamination from the NOT spectra has been removed following the 
  prescriptions
  of \protect\cite{stan07}. The noisy spectrum obtained on 22 February, 2009, was slightly smoothed.
  The apparent narrow P-Cygni feature visible at the  
  position of He I $\lambda$5876 in the 2009 January 22 spectrum is in reality a spike due to a poorly removed hot pixel. Some spectra  
  have been cut at the blue wavelengths because of the modest  S/N. 
} \label{fig7} 
\end{figure*}

\begin{figure} 
\includegraphics[width=90mm]{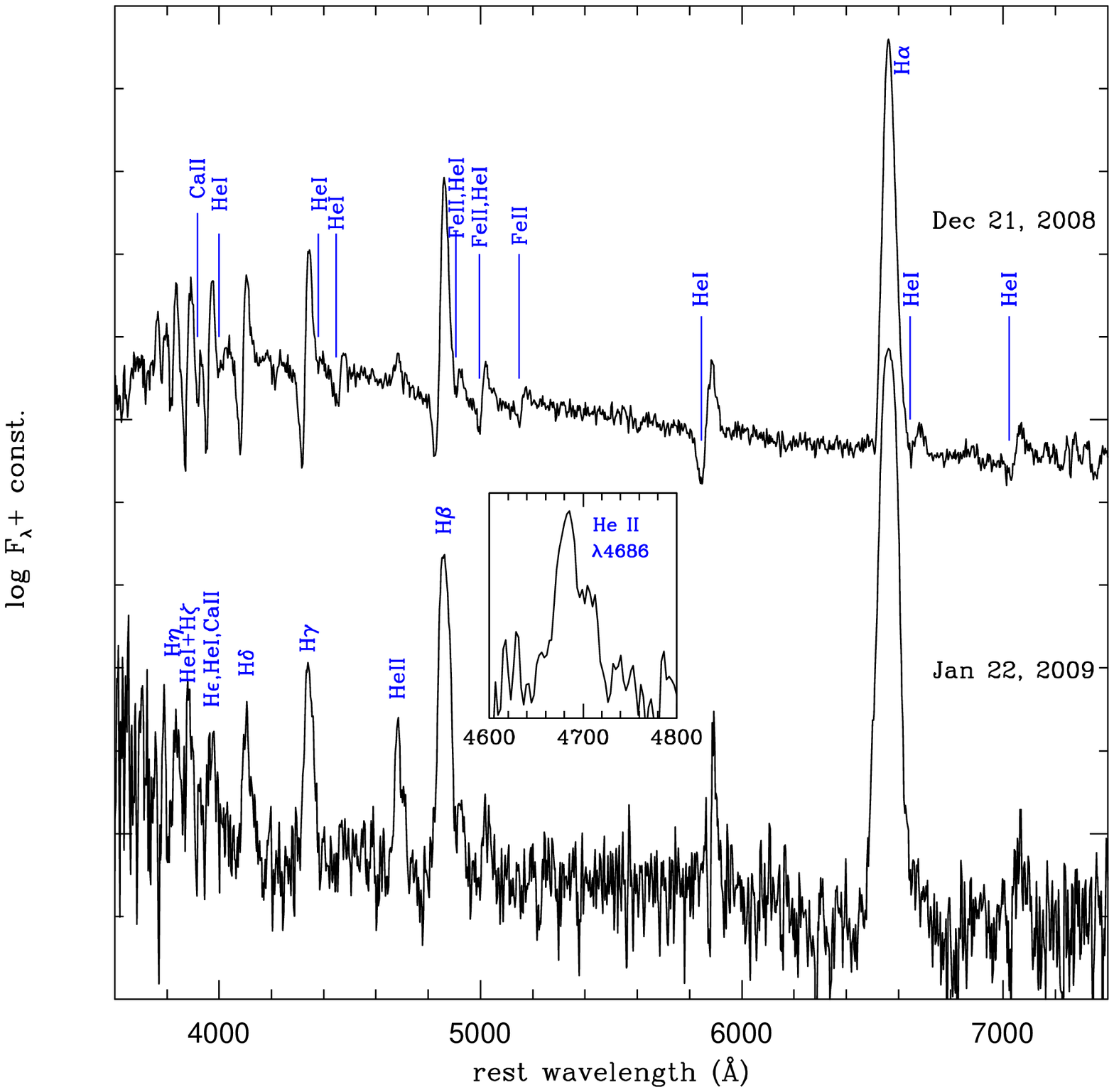} 
  \caption{Line identification in the spectra obtained on December 21, 2008, and January 22, 2009.  
  The apparent narrow P-Cygni feature visible at the  
  position of He I $\lambda$5876 in the 2009 January 22 spectrum is in reality a spike due to a poorly removed hot pixel.  The
  insert shows in detail the complex structure of the He II $\lambda$4686 line in the January 22 spectrum.
} \label{fig8} 
\end{figure}

Spectra of the luminous variable in NGC 3432 were obtained after the October 2008 outburst using the 4.2-m William Herschel Telescope (WHT) equipped with ISIS, the 3.58-m Telescopio Nazionale 
Galileo (TNG) with LRS, the 2.56-m Nordic Optical Telescope (NOT) with ALFOSC and the 2.2-m Telescope of the Calar Alto Observatory plus CAFOS.  
Information on the spectra obtained during the period from October 2008 to November 2009 is reported in Table \ref{tab3}. 
The best signal-to-noise (S/N) spectra of the star witnessing the evolution of the 2008-OT and 2009-OT1 outbursts, but also the 
subsequent quiescent phases, are shown in  
Figure \ref{fig7} together with spectra obtained in 2000/2001 (2000-OT) from \citet{wag04}.  
The spectra of the variable obtained soon after 2008-OT are characterized by prominent H lines in emission (see line identifications in Figure \ref{fig8}).  
He I lines are also clearly visible with $v_{FWHM} \approx$ 2000 km s$^{-1}$, which is only slightly lower than that of H$\alpha$ 
($v_{FWHM} \approx$ 2300 km s$^{-1}$). The spectra of the 2008 episode  display  
significant differences compared with those of 2000-OT described in \citet[][and shown atop of Figure \ref{fig7}]{wag04}. In the 2000 event the He I lines were  
weaker and the FWHM velocities measured for H$\alpha$ were slightly lower, being in the range 1600-1800 km s$^{-1}$. Such differences in line
velocities may suggest a more powerful outburst (but we cannot confirm it on the basis of our photometric data) or less envelope mass
ejection in the 2008 outburst. However, the integrated flux of H$\alpha$ obtained  
averaging the flux measured from the three available spectra of October 2008 is about 4 $\times$ 10$^{-14}$ erg cm$^{-2}$ s$^{-1}$, which is a factor of two more than   
the flux reported by \citet{wag04} for the 2000 event, implying that the 2008 event had an H$\alpha$ luminosity twice that of the 2000 event.  
Adopting the same relation between the H$\alpha$ luminosity and the radius (R) of the emitting region and the same assumptions as Wagner et al., we  
obtain $R(H\alpha)$ = 0.25 pc \citep[from the H$\alpha$ luminosity computed from the 2000 May 14 spectrum, $R(H\alpha)$ = 0.2 pc was determined by][]{wag04}.  
Remarkably, the peak of the H$\alpha$ line is asymmetric and slightly blue-shifted. This phenomenon is visible also in the late-time spectra  
of some core-collapse SNe and is expected when newly-formed dust extinguishes the light coming from the receding hemisphere  
\citep[see e.g. SN 1999em,][]{elm03}. 
  
High S/N spectra were then obtained on December 21-22, 2008, showing a remarkable evolution. The continuum 
is very blue, and prominent P-Cygni profiles are now visible. It is worth noting that P-Cygni profiles were marginally detected (especially in H$\beta$)  
also in the 2000-OT spectrum of May 6, 2000 \citep{wag04}. 
Together with H and He I, Fe II lines of the 
multiplet 42 and Ca II H$\&$K\footnote{The identification of Ca II is unequivocal, being the line at $\lambda$ 3934\AA~unblended 
in the Nordic Optical Telescope spectrum of December 21st, 2008.} are clearly detected 
(see Figures \ref{fig7} and \ref{fig8}). Now H$\alpha$ does not show a P-Cygni profile and its $v_{FWHM}$ is about 2000 km s$^{-1}$, consistent with the velocity deduced 
from the position of the minimum of H$\beta$. Other Balmer lines show significantly lower
P-Cygni velocities (about 1600 km s$^{-1}$),  
very close to those of  He I lines ($\sim$1500-1550 km s$^{-1}$). Finally, unblended Ca II $\lambda$ 3934 and Fe II $\lambda$ 5169 
P-Cygni lines have absorption components which are blue-shifted by about 1250 km s$^{-1}$. Comprehensive line identification is shown in Figure 
\ref{fig8}. 
 
A further spectrum of the variable was obtained on January 22, 2009. Within one month, the spectrum of the variable has evolved significantly. 
P-Cygni profiles are no longer visible, He I and Fe II lines are marginally detected, while a prominent He II $\lambda$ 4686 line is now visible 
in emission (see insert in Figure \ref{fig8}). The same line was visible also in the second spectrum of 2000-OT (May 31, 2000) presented by \citet{wag04} and, marginally, 
also in the January 19, 2001, spectrum. The presence of this line would  
indicate high temperatures of the emitting material. 
He II lines in emission are common features in He-rich, hot Wolf-Rayet stars and early O-type supergiants,
and are only rarely detected in LBVs during quiescence \citep[e.g. in AG Car,][]{sta86}. 
Finally, two spectra were collected on 2009 February 22 and March 22, and the object did 
not show any further evolution. We noted only a weakening of the He II line. The line widths are very similar
to those measured in the January spectrum, viz. $v_{FWHM} \approx$ 2750 km s$^{-1}$. 

\begin{figure} 
\includegraphics[width=90mm]{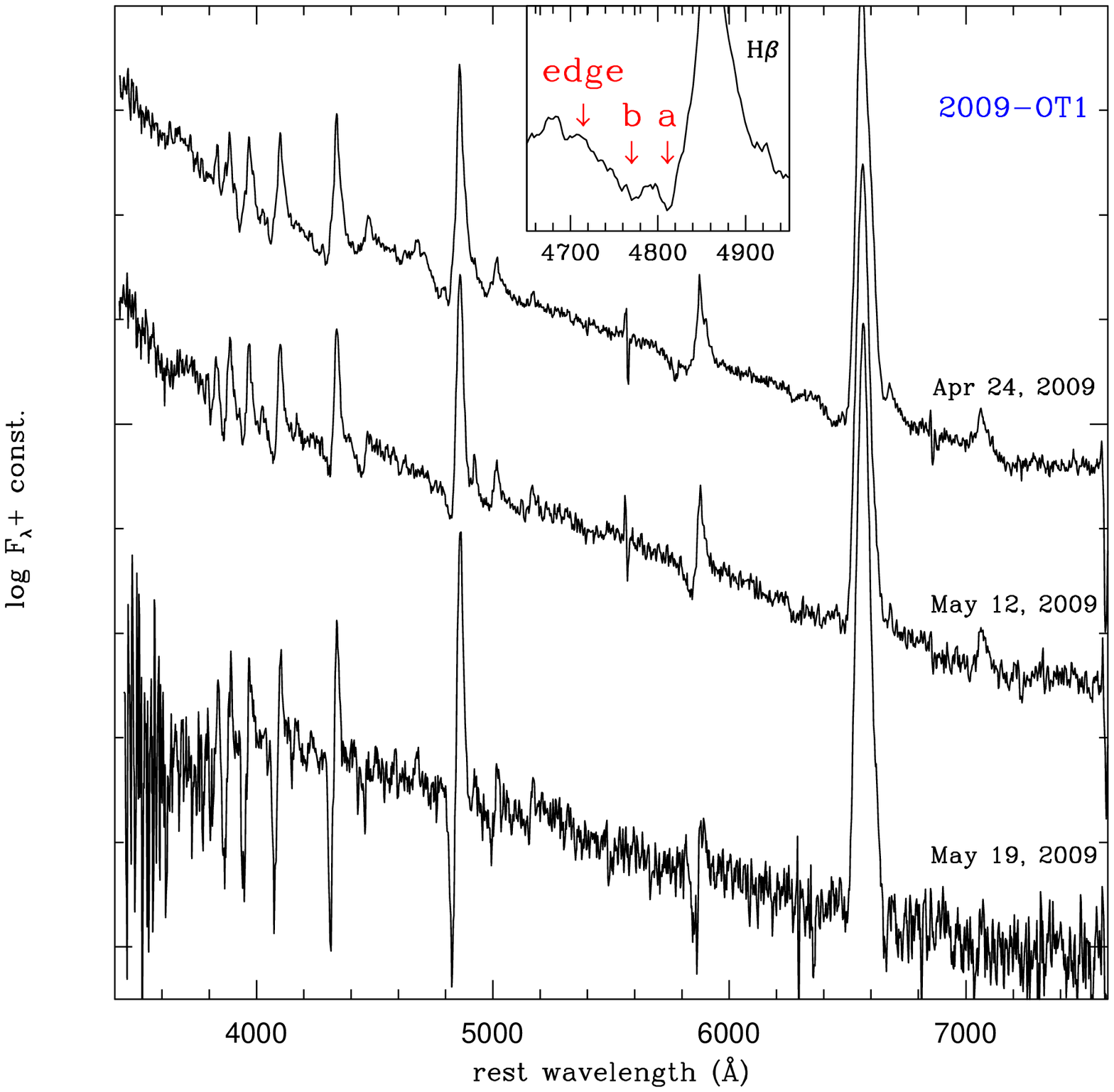} 
  \caption{Spectroscopic evolution of the April-May 2009 outburst (2009-OT). A zoom-in of the H$\beta$ region of the April 24 spectrum 
is shown in the insert. Individual features mentioned in the main text are labelled.  
} \label{fig9} 
\end{figure}

The spectra obtained during the 2009-OT1  event (April-May 2009, see Figure \ref{fig9}) show prominent P-Cygni lines of H, He I and Fe II,  and
share striking similarities with the December 2008 spectra, although the continuum is now significantly bluer. This is probably due to the fact 
that these spectra were obtained closer to the epoch of the outburst onset. From May 12th  
to May 19th the continuum temperature decreases significantly and the spectral lines show more prominent P-Cygni absorption components. 
The H and He I lines in the first spectrum of Figure \ref{fig9} (April 24, 2009)  
show two-component absorption profiles. We identify a lower-velocity trough (labelled with {\sl a}~
in Figure \ref{fig9})  
which is blue-shifted by about $v_a \approx$ 3000 km s$^{-1}$, and another absorption component (labelled {\sl b}) with a core velocity $v_b$  
of about 5300 km s$^{-1}$. The broad wings of the absorption feature extend up to a velocity of $v_{edge} \approx$ 9000 km s$^{-1}$.
As H$\beta$ is not saturated, $v_{edge}$  is not a perfect indicator for the terminal wind velocity ($v_\infty$).
\citet{pri90}  found that $v_{edge}$ overestimates $v_\infty$ by 20-25$\%$ ($v_\infty \approx$ 0.8 $v_{edge}$), and we can therefore
more accurately estimate the terminal velocity of H$\beta$ as 7200 km s$^{-1}$.

This peculiar line profile is remarkably similar to those observed by \cite{tru08} in SN 2005gj.  
The  similarity of this absorption line splitting in the SN spectra with those of well known LBVs \citep{sta01,sta03} led \cite{tru08} to conclude that the progenitor of SN 2005gj 
was likely an LBV that experienced a mass-loss episode a few decades before the SN explosion.  
In the spectrum of 2009-OT1 we measure velocities which are more than one order of magnitude higher that those reported by \cite{tru08}, very
different to those of ordinary LBV winds. 
The higher velocity trough (b) is only marginally detectable in the subsequent spectrum of 2009-OT1 (May 12, 2009), while the lower velocity one is  
still  visible and has $v_a$ $\approx$ 2000-2200 km s$^{-1}$. The whole absorption feature extends to blue wavelengths corresponding to   
$v_{edge} \approx$ 6000-7000 km s$^{-1}$ ($v_\infty \approx$ 4800-5600 km s$^{-1}$). The last spectrum of the variable in Figure \ref{fig9} (May 19, 2009) shows P-Cygni lines 
with a single absorption component, though with a slightly asymmetric profile.  
The minimum of the absorption indicates $v_a \sim$ 2000 km s$^{-1}$, but with the blue wing extending to $v_{edge} \approx$ 3000-3500 km s$^{-1}$ ($v_\infty \approx$ 2400-2800 km s$^{-1}$). 
 
The last spectrum of the variable in NGC 3432 (see Table \ref{tab3}) was  obtained at the time of the 2009-OT2 outburst 
(November 26, 2009), but it is not shown in Figure \ref{fig7} because of its poor S/N. 
It showed a blue continuum and narrow H lines, with $v_{FWHM}$ of about 1300-1400 km s$^{-1}$,
still consistent with the lower line velocities displayed by this variable during all earlier outbursts.

\citet{vink02} suggested that oscillations in the wind velocity (and mass-loss rate) similar to those we observe in
the spectra of the variable in NGC 3432 can be associated with changes in the
efficiency of line driving, as a result of variations in the Fe ionization of the
wind \citep[bi-stability in the wind velocity in early-type stars, see][]{lam95,vink99}. We note that 
also \citet{koe06} invoke the bi-stability mechanism to explain the changes in wind velocity of HD 5980 (see Sect. \ref{prog}).
Despite the high degree of variability, the wind velocities of the variable in NGC 3432
remain remarkably high both during outbursts and in quiescence, making an ordinary LBV 
scenario for this object rather problematic.
 
\subsection{Comparison with Luminous Stars and Stellar Outbursts}
 
\begin{figure*} 
\includegraphics[width=81mm]{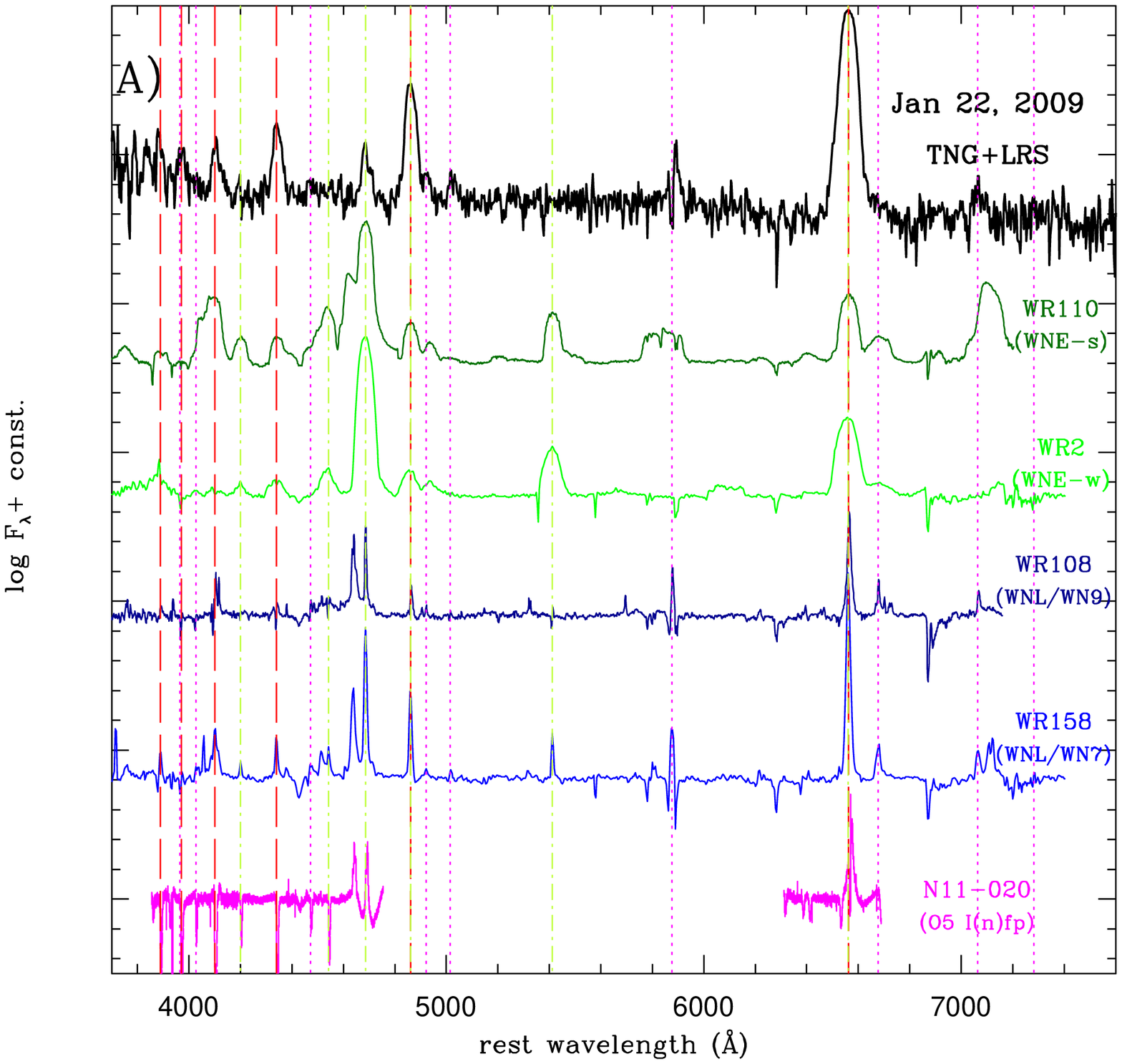} 
\includegraphics[width=81mm]{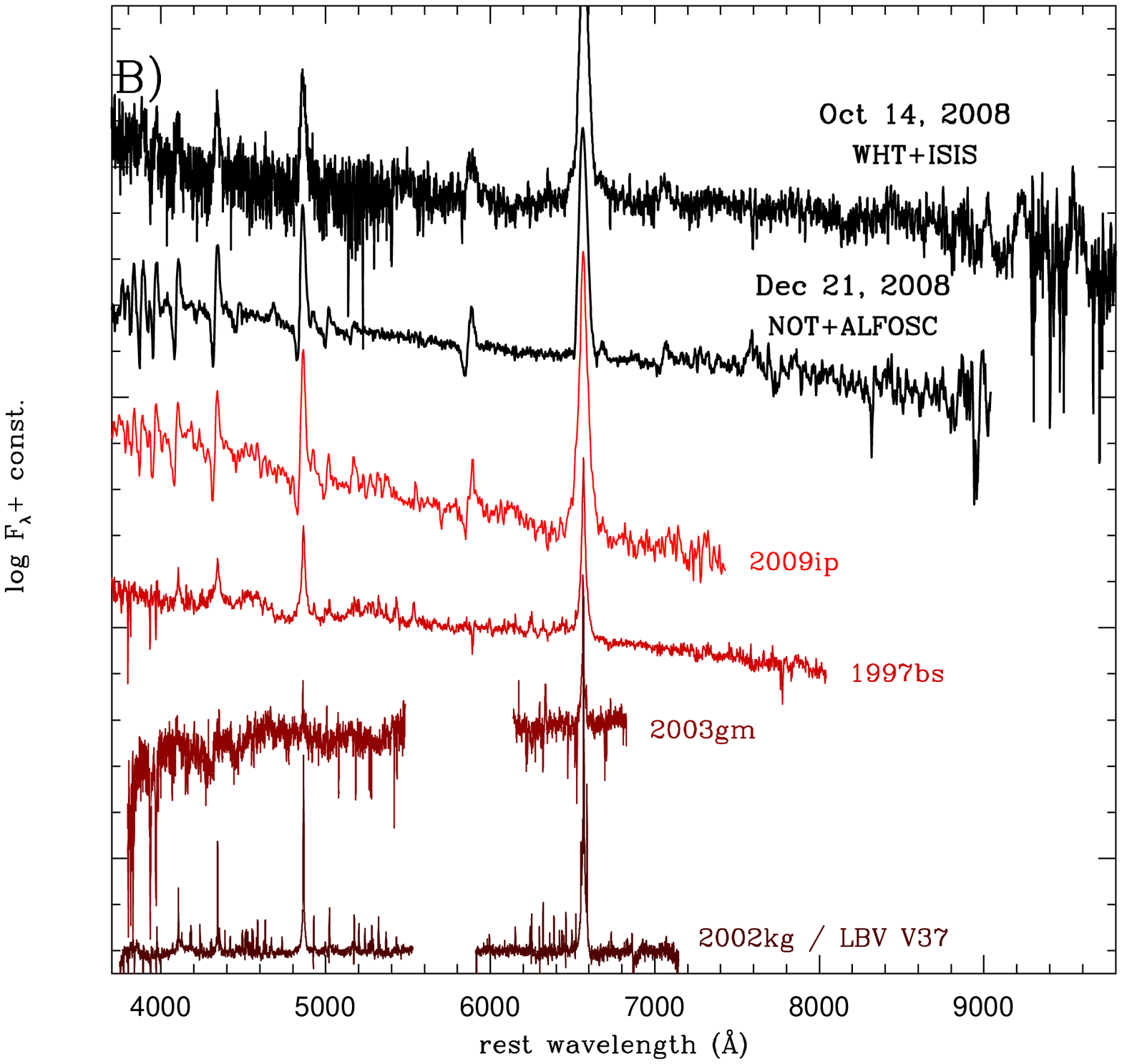} 
\includegraphics[width=81mm]{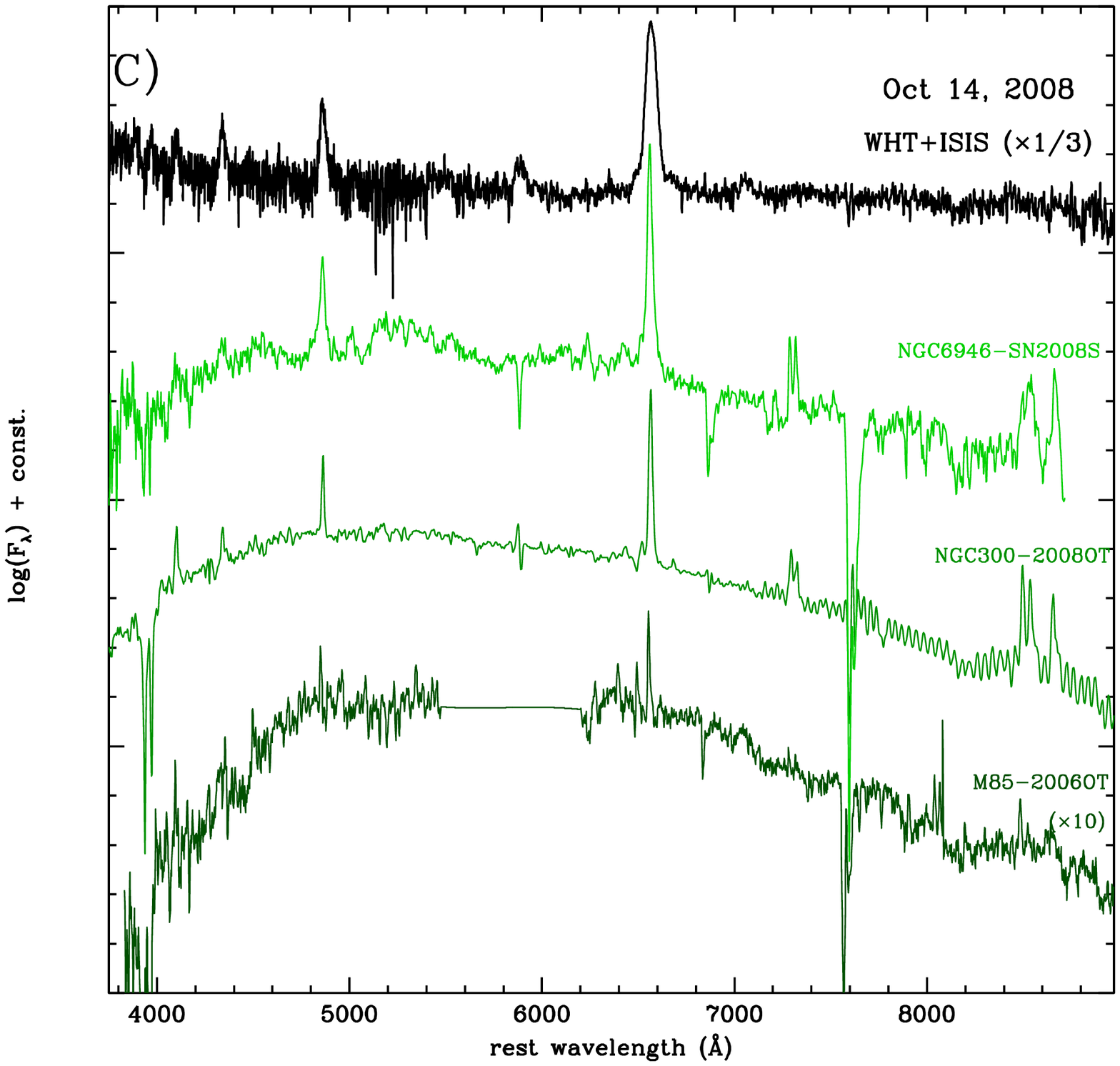} 
\includegraphics[width=81mm]{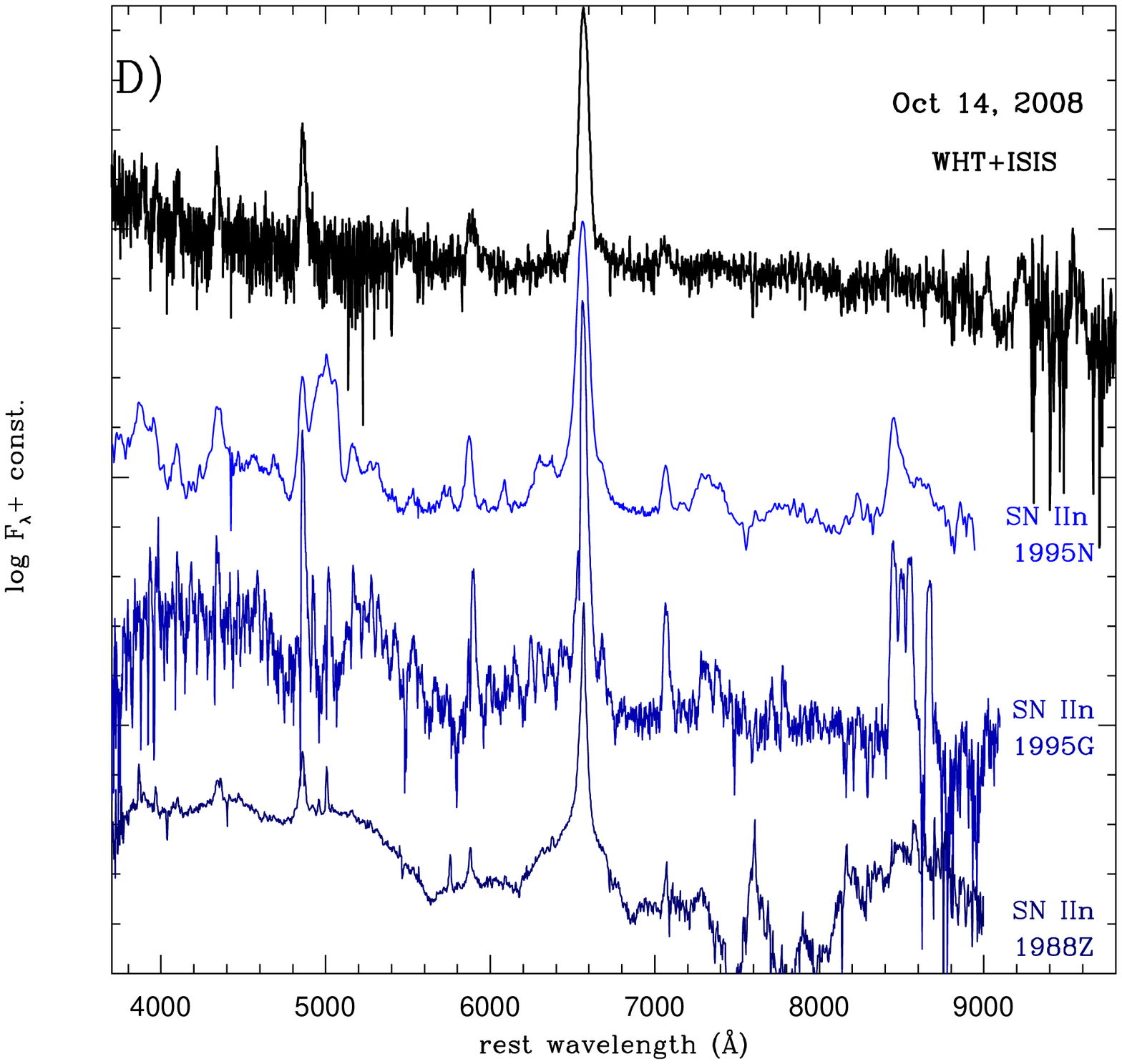} 
  \caption{Comparison of spectra of the variable in NGC 3432 with those of different families of objects. 
  {\bf A.} Comparison of a spectrum of the variable in quiescence (January 22, 2009) 
  with those of a few Wolf-Rayet (WN) stars and an Oe-type star: WR 110
  \protect\citep[WNE with strong lines; WN6-s, see][]{ham95a}, WR 2 (WNE with weak emission lines; WN2-w), WR 108 (WNL; WN9), 
  WR 158 (WNL, WN7), and the O5-type N11-020. The positions of the most important H lines are marked with 
  red dashed lines, those of He I features are indicated with magenta dotted lines  
  and those of He II with bright green dot-dashed lines. 
  Note that WR 2 and WR 110 do not show clear signature of H lines, and the features 
  at the positions of H$\alpha$ and H$\beta$ are mostly due to He II. All spectra of WN stars shown in the figure  
  are from \protect\citet{ham95b}, that of N11-020 from \protect\cite{evan06}. The narrow spike at the position of He I 5876\AA~ in the spectrum of 2008-OT 
  is due to a poorly-removed hot pixel. 
  {\bf B.} Comparison of two spectra of the variable in NGC 3432 with those of other well-known SN impostors: 
  2009ip (ESO-NTT spectrum obstained on October 22, 2009; Padova-Asiago Supernnova Archive), 1997bs \protect\citep{van00},  2003gm and 2002kg \protect\citep{mau06}. All these transients 
  are thought to be luminous eruptions of LBVs.  
  {\bf C.} Comparison of the spectrum of the variable in NGC 3432 after the 2008 outburst (October 2008, 14)  
  with those of a sample of enigmatic transients whose nature is still debated (see text). The spectrum of 
  (SN) 2008S is from \protect\citet{bot08}, that of the optical transient in NGC 300 is from Pastorello et al. 
  (in preparation), and that of M85 OT 2006-1 is from \protect\citet{kul07}. 
  {\bf D.} Comparison of the 14th October  
  spectrum of the variable in NGC 3432 with those of luminous, interacting core-collapse SNe: the type IIn SNe 1995G, 
  1995N and 1988Z \protect\citep[from][respectively]{pasto02,pasto05,tura93}. 
} \label{fig10} 
\end{figure*}

During the LBV phase, massive stars vary
 their apparent temperatures and, as a consequence, spectral types.
The high degree of variability of LBVs is generally evident in their spectra.
During the quiescent phase LBVs tend to have blue spectra
often similar to those  of B supergiants. However, a number of LBVs such as AG Car and HD 269582 \citep{sta86}
show Opfe/WN9 spectra during quiescence. In this phase, these objects have also strong He II in
emission, that disappears during outburst when the spectra display prominent P-Cygni profiles.
Nevertheless, the remarkably high velocities measured in the spectra of the variable in NGC 3432 both during outburst
and in quiescence are unusual. 

 In order to better constrain the nature of this variable, we compare its spectra with those 
of well-known classes of luminous stars or other types of transients. 
In Figure \ref{fig10} (panel {\bf A}) a spectrum of the variable star during the quiescence after 2008-OT is shown along with those of a number of  Wolf-Rayet 
(of WN type) stars. While early WN (WNE) stars show relatively broad emission lines with a FWHM which is quite consistent with 
that observed in the spectrum of the variable in NGC 3432, their spectra are characterized by prominent He II emission lines, which dominate over the H features. 
(Unblended He II line at  4686\AA\ is visible only in the 2009 January 22 spectrum, when the star was quiescent, and its intensity 
was much lower than that observed in spectra of WNE stars).  
Late WN (WNL) stars have instead more intense H lines, but very narrow. In addition, lines such as N III, N IV, C III, C IV which are  
prominent in WN stars, are not visible in the spectra of the variable \citep[see][for a detailed information on the sub-classification of WN stars]{ham95a,cro95,cro95b,cro95c,nota96}. 
The spectrum of the variable in NGC 3432 shares some similarity with those of other luminous hot stellar types, such as Be, Oe stars and O-type supergiants. Although Be stars show 
conspicuous H$\alpha$ lines in emission, sometimes the line profiles are double-peaked, suggesting the presence of a disk
\footnote{Note, however, that only linear spectropolarimetry can unveil the real geometry of a stellar system. \protect\citet{vink09}, 
indeed, recently questioned the disk-hypothesis for Oe stars due to the lack of polarimetric proofs.}. 

Finally, Oe and O-supergiant stars 
still show prominent H lines, and He II features are weak (usually weaker than He I lines). This is the case of N11-020, classified as O5 I(n)fp by  
\citet[][see Figure \ref{fig10}, bottom of panel {\bf A}]{evan06}, whose spectral lines are relatively broad, but still too narrow in comparison with those of the variable.  
In addition, lines of other ions visible in all 
stellar types described above (N II, N III, C III, C IV, Si IV) {\it are not visible} in the spectra collected so far  
for the luminous star in NGC 3432. 

In Figure \ref{fig10} (panel {\bf B}) two spectra of the variable in NGC 3432 are compared with those of the SN impostors 1997bs \citep{van00}, 2009ip (Padova-Asiago Supernova Archive), 2003gm \citep{mau06} and  
2002kg \citep{mau06}. 
It is worth noting that the spectral lines of the NGC 3432 variable are  broader than those of the other SN impostors. H$\alpha$ is  
indeed dominated by an intermediate-width component ($v_{FWHM} \approx$ 2300 km s$^{-1}$) and weak,  
marginally detectable broader wings.

 The other SN impostors shown in Figure \ref{fig10} have FWHM velocities of H$\alpha$ that typically are $\leq$ 1000 km s$^{-1}$, similar 
to that measured for 2001ac \citep[$v_{FWHM} \approx$ 900 km s$^{-1}$,][]{mat01}. Only 2009ip competes with the variable in NGC 3432 in line widths \citep[see also the discussion
in][]{smi09b,fol10}, although its spectrum does not show He II $\lambda$4686. 
The spectrum of 1997bs 
presented by \citet{van00} and shown in Figure \ref{fig10} (panel {\bf B}) has a two-component emission profile, with a narrow component ($v_{FWHM} \approx$ 500 km s$^{-1}$)  
atop a broader one, with  $v_{FWHM} \approx$ 1500 km s$^{-1}$. 
In this context, the spectral lines of the major outbursts of the variable in NGC 3432 are 
broader by a factor of 1.5-3, with FWHM velocities which are close to those observed in the intermediate components of some  
interacting SNe (see Figure \ref{fig10}, panel {\bf D}). Nevertheless, the most important spectral features 
observed in the variable in NGC 3432 (due to H, He I and Fe II transitions) are commonly observed also in the spectra of SN impostors \citep[see 
e.g.][]{mau06}. The only exception is the He II $\lambda$ 4686 line visible in the quiescent spectrum of the variable, which is rarely observed 
in spectra of erupting LBVs. 

We compare in Figure \ref{fig10} (panel {\bf C}) the 2008 October 14 spectrum of the NGC 3432 variable with those of a sample of enigmatic transients whose 
overall similarity was first discussed by \citet{pri08b} and \citet{tho08}. The nature of this group of transients, which includes  
SN 2008S \citep{pri08a,bot08,smi08}, NGC 300 2008-OT \citep{bond09,ber09} and  
M85 OT 2006-1 \citep{kul07}, is still debated. According to \citet{smi08}, \citet{bond09} and \citet{ber09} these transients 
are likely SN impostors caused by the luminous eruptions of massive (10-20 M$_\odot$) stars.  
While the progenitors of SN 2008S and NGC 300-OT were identified as dust-enshrouded massive stars in archive {\it Spitzer} images \citep[of about  
10M$_\odot$, according to][]{pri08a,tho08,bot08}, 
the star producing M85 OT 2006-1 was never detected, and the outburst was associated to the merging of low-mass stars by \citet{kul07}, \citet{rau07}
and \citet{ofek08}.  
\citet{pasto07b} questioned both the mass limit derived for the progenitor of M85 OT 2006-1 and the nature of the outburst itself, proposing that it was 
the underluminous core collapse of a $\sim$8M$_\odot$ star. Finally, \citet{bot08} discussed the possibility that SN 2008S and probably also the other transients of this 
group were electron-capture SNe from super-AGB stars \citep[see also][]{wan08,pumo09}.  
H, He I and Fe II lines are detected both in the spectra of the luminous star in NGC 3432 and in objects of the family.  
However, the latter objects show a prominent [Ca II] feature at about 7300\AA~ (not visible in our spectra  
of the NGC 3432 variable), narrower spectral lines  
and a much slower photometric evolution. 

Our first spectrum after the October 2008 outburst is finally compared in Figure \ref{fig10} (panel {\bf D}) with those of three type IIn SNe \citep[1995G, 1995N, 1988Z; 
see][]{pasto02,pasto05,tura93}. The spectra of SNe 1995G and 1995N were obtained more than one year  
after their explosions, while that of SN 1988Z is probably earlier. The FWHM of the lines in the spectrum of the variable in NGC 3432 is similar to that of the  
intermediate-velocity line components of the three type IIn SNe \citep[see][]{fra02,pasto02,zam05,tura93,are99}. In interacting SNe these line components are thought to originate from
the shocked CSM.

All these comparisons indicate that the spectra alone do not allow us to discriminate between the  
different types of explosions. Additional information provided by the photometry (specifically the low peak luminosity, the fast  
luminosity evolution of the outburst, the variability history of the star) makes us confident in identifying the 2000/2008/2009 transients in NGC 3432  
as the most recent episodes of a long series of major mass-loss events in this exceptional, restless variable.  
       
\begin{figure*} 
\includegraphics[width=170mm]{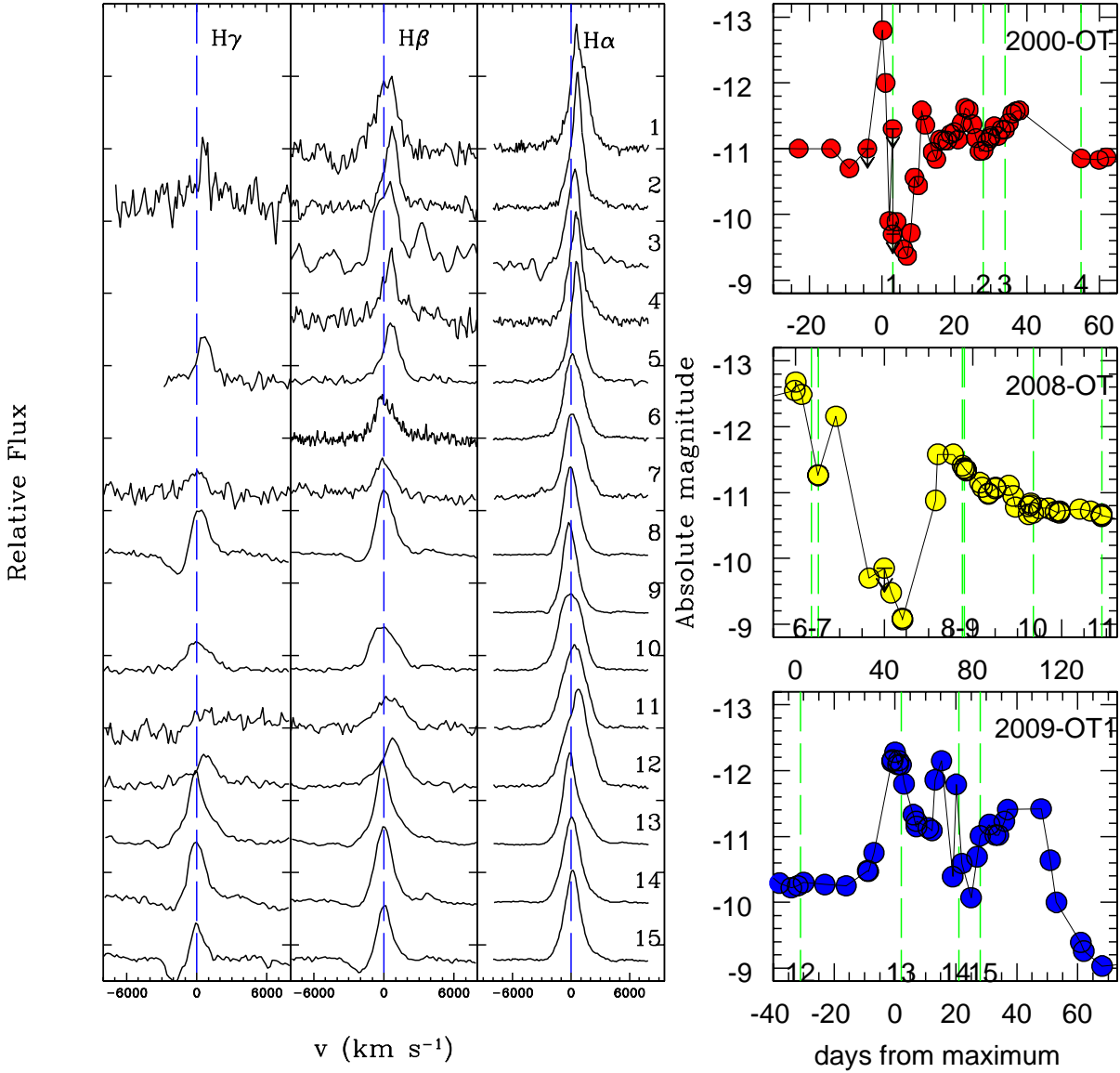} 
  \caption{The profiles of H$\gamma$, H$\beta$ and H$\alpha$ in the spectra of the luminous variable in NGC 3432 between 2000 and 2009 (left panel). 
  The numbers in the panel of H$\alpha$ allow us to date the spectra in the light curves of the three outbursts
shown in the panels on the right: 2000-OT (top), 2008-OT (middle), 2009-OT1 (bottom). The vertical green dashed lines 
  mark the epochs at which spectra are available: May 6th, 2000 (1); May 31st, 2000 (2); June 5th, 2000 (3); June 27th, 2000 (4);
  January 19th, 2001 (5); October 14th, 2008 (6); Oct 17th, 2008 (7); December 21st, 2008 (8); December 22nd, 2008 (9);
  January 22nd, 2009 (10); February 22nd, 2009 (11); March 22nd, 2009 (12); April 24th, 2009 (13); May 12th, 2009 (14); May 19th, 2009 (15). 
  Some spectra have been cut at the blue wavelengths because of the poor 
  S/N. The epoch of the January 19, 2001 spectrum  \protect\citep[labelled with number 5,][]{wag04} is not marked on  
  the right-hand panels because the spectrum  was obtained several months after 2000-OT. The low S/N spectrum of June 5, 2000 \protect\citep{wag04} was slightly smoothed.} 
  \label{fig11} 
\end{figure*} 

\subsection{Evolution of the Line Profiles} \label{prof} 
 
A view of the spectral sequence in Figure \ref{fig7} suggests that two types of spectra were observed during
follow-up observations of the variable in NGC 3432 (October 2008 to June 2009).  
\begin{enumerate} 
\item A few spectra (marked with 1, 8, 9, 13, 14, 15 in Figure \ref{fig11}) show a blue continuum with prominent P-Cygni lines of H, He I and (although weak) Fe II. On average, the FWHM 
velocity of the emission features is around 1600-2200 km s$^{-1}$. In these spectra there is no clear evidence for the presence of He II lines. 
The reason why the spectral energy distribution (SED) peaks at the blue wavelengths during outbursts will be addressed more in detail in Section \ref{sed}.
\item Other spectra (e.g. 2, 3, 4, 10, 11, 12) show a much redder continuum and lines of H and He I in pure emission. The FWHM velocities are significantly higher 
(2400-2800 km s$^{-1}$) \footnote{In the spectra of the variable obtained during the period 2000-2001, the FWHM velocities were  
slightly lower,  between 1600 and 2100 km s$^{-1}$ in the post-outburst phases.}. In these spectra, the He II  $\lambda$4686 feature is unequivocally detected. 
\end{enumerate} 

In order to understand the reasons of the rapid variability of the spectra, we have visualized  
in Figure \ref{fig11} the evolution of the profile of the most important H lines, and correlated  
the spectra with the phases of the three outbursts. 
In particular, for each spectrum in Figure \ref{fig11} (left panels) we have assigned an identification number. The phases of the light curves  
reported in the right panels of Figure \ref{fig11} are relative to the epoch of the outburst peaks. We have marked  
the epochs of the individual spectra with vertical  
dashed lines. We note that the low S/N spectrum marked with 1 (May 6, 2000), which shows 
evidence of H$\beta$ with P-Cygni profile, was captured when the object was in the deep, post-maximum decline (R $>$ 20.5)
but still very close in time ($\sim$3 days) to the epoch of the outburst. 
Other spectra obtained soon after the outburst episodes show an excess at blue wavelengths 
and  clear P-Cygni profiles. On the other hand, spectra with pure emission lines, including
He II 4868\AA, were mostly obtained in periods of relative quiescence (R between 19 and 20). 
In addition, counter intuitively,
we found that $v_{FWHM}$ of the line components in emission is lower immediately after outbursts than when the star is quiescent. 

In other words, close to the outburst epochs, the spectra of the variable in NGC 3432 are reminiscent to those
of regular erupting LBVs, while in quiescence the spectrum acquires some resemblance with those of young WRs.
As mentioned above, although spectra of some LBVs may show He II lines during  quiescence 
\citep[e.g. AG Car,][]{sta86}, the velocity of the  material ejected by the variable in NGC 3432 (as deduced from spectroscopy)
is too high for a single LBV eruptor.
As an alternative, supported by marginal evidence of modulation in the light curve,
one may propose a multiple-system scenario. During outburst the total flux is dominated by
the LBV eruptor, while in quiescence we start to see a WR companion.
This scenario, similar to that proposed for the famous stellar system HD 5980 \citep[see e.g.][and references therein]{koe04},
is widely discussed in Section \ref{prog}.
 
\section{Spectral Energy Distribution} \label{sed} 
 
 \begin{figure*} 
\includegraphics[width=132mm,angle=270]{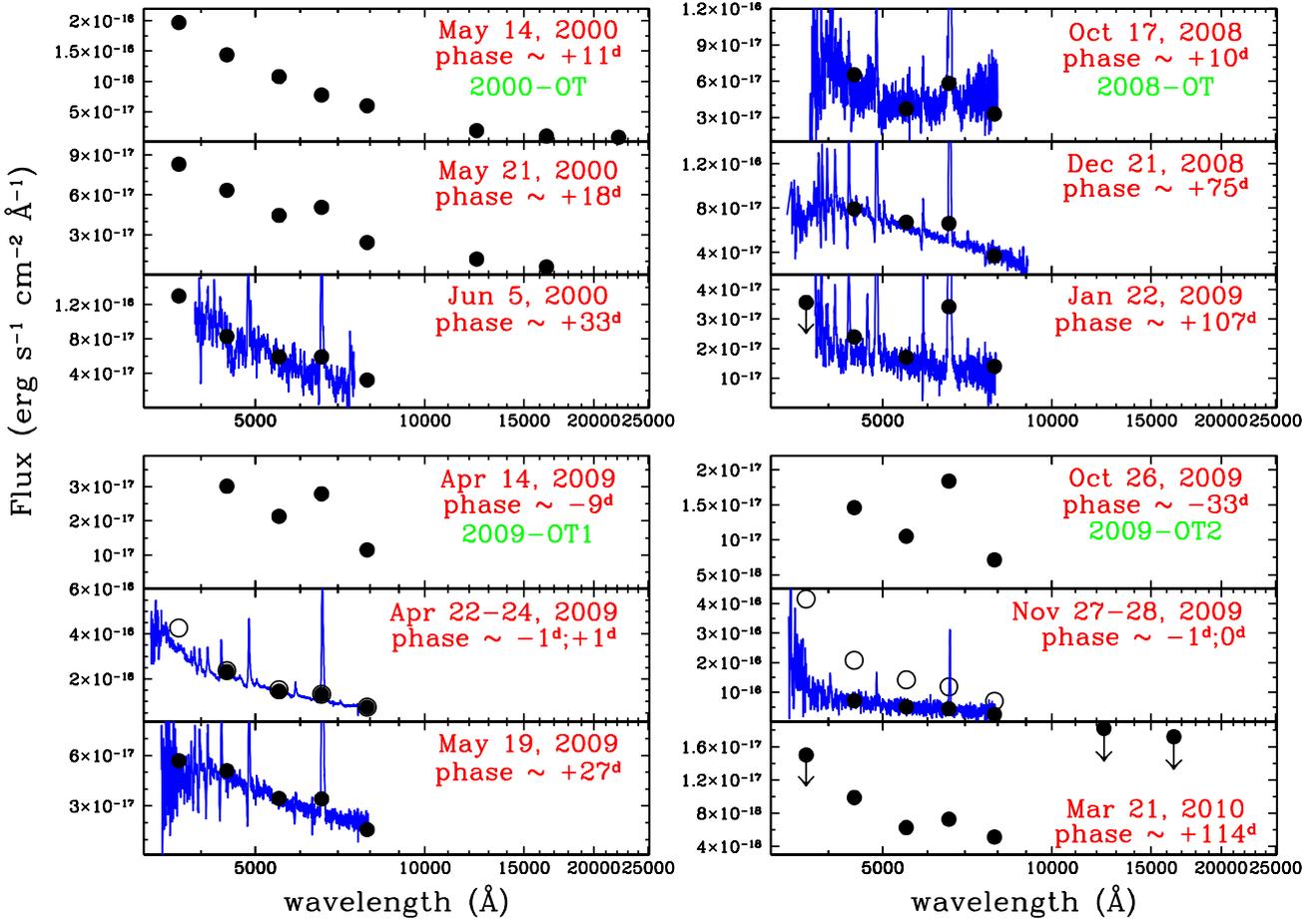} 
  \caption{Evolution of the SED of the variable in NGC 3432 at 12 representative epochs (labelled 
  on the right). Coeval spectra have also been shown as a comparison at 7 epochs.  
The spectrum and the SED of April 24th, 2009 (filled circles) are shown together with the SED 
  computed 1 day before the maximum of 2009-OT1 (April 22nd, 2009; open circles). Analogously, the spectrum and the SED of November 27th, 2009 (filled circles)
  have been shown together with the SED computed at the epoch of the peak of 2009-OT1 (November 28th, 2009; open circles).} 
  \label{fig12} 
\end{figure*} 

The evolution of the SED for the variable in NGC 3432 is shown in Figure \ref{fig12}.
The four panels show the SED at 3 epochs representative of crucial phases in the evolution of each major outburst:
2000-OT (top-left panel), 2008-OT (top-right), 2009-OT1 (bottom-left) and 2009-OT2 (bottom-right). 
We account only for epochs in which multi-band observations of the source are available and,
typically, close to a light curve maximum, a minimum and in quiescence. 

As mentioned in Section \ref{ph}, the light curve peak and the subsequent minimum after 2000-OT are extremely sharp, 
and multi-band observations are available only from day +11 \citep{wag04}, when the transient is in the post-minimum
rise. The SEDs at this and subsequent epochs (phases $\sim$ +18 and +33; Figure \ref{fig12}, top-left panel)
are extremely blue, with a strong U-band flux contribution (the reddening-corrected U-B ranges from -0.6 to -0.8 during this period). 

The evolution of the SED after the 2008-OT eruptive episode is shown in the top-right panel. Again, multi-band 
observations of the luminosity peak are missing. However, at day +10 from the main outburst (in coincidence with 
a sharp minimum) the source shows a redder SED ($V-I \approx$ 1), with a clear R-band excess probably due to the flux contribution of 
H$\alpha$. The SED is then computed at +75 days when it becomes again much bluer
($U-B \approx$ -0.6; $V-I \approx$ 0.5), which
is probably the result of a secondary eruptive event\footnote{There is a rebrightening in the light curve at day $\sim$60 after the peak of 2008-OT, clearly visible e.g. in 
Figure \ref{fig11}, middle-right panel.}. 
Finally, at +107 days from the main event, the absolute
magnitude of the object is about $M_R \approx$ -10.7, close to that expected during the quiescence. In this phase, there is
a huge R-band (H$\alpha$) excess and the colour is still very red ($V-I \approx$ 1).

In Figure \ref{fig12} (bottom-left panel, upper insert) we report the SED of the star at $\sim$9 days before the peak of
the 2009-OT1 outburst, when the luminosity of the object is rising after a deep minimum, and is close to the magnitude at the quiescence.
The SED is quite similar to that computed on 2008 October 17. The central insert shows the SED around maximum light,
which is extremely blue ($U-B \approx$ -1, $V-I \approx$ 0.4). In the subsequent epoch (+27 days) the SED is computed during a sharp luminosity 
minimum, but still with the object being in a very active, eruptive phase. Consequently, despite the faint absolute magnitude 
($M_R \approx$ -10.3), the variable still showed a moderately blue SED ($U-B \approx$ -0.5, $V-I \approx$ 0.4).

The last panel in Figure \ref{fig12} (bottom-right) shows the SED evolution before (-33 days, in a deep magnitude 
minimum, with $M_R \approx$ -9), during and after (+114 days) the 2009-OT2 outburst. The evolution of the SED is similar to that  shown
for 2009-OT1 (bottom-left panel). At day -33 there is still a huge R-band excess and the SED is very red ($V-I \approx$ 0.8),
then it becomes blue during outburst ($U-B \approx$ -0.9, $V-I \approx$ 0.45) and finally redder again in quiescence
($V-I \approx$ 1 at day +114). Unfortunately, the U, J and H band limits reported at phase $\sim$ +114 days are not stringent.

 The observed SED is dominated by the U-band emission at all epochs that are relatively close in time to an eruptive episode. 
 This is clearly visible around and soon after the 2009 episodes (2009 April 22-24 and November 27-28), but also on 2008 December 21 when possibly a secondary eruptive episode
 occurred after 2008-OT. In our attempts to fit the SED around the
 time of the outbursts with a single blackbody we obtain reasonable fits to the fluxes in the B, V, R, I bands with temperatures
 of 8500-9500 K, whilst we fail to simultaneously fit the U band flux, which shows a clear excess.
 At these epochs, the spectra show prominent and narrow P-Cygni features, and there is no evidence for the presence of He II $\lambda$4686.
 After the outbursts, the SED evolution suggests a decline of the temperatures \citep[in agreement with][who fitted the optical SED of 2000 May 14, i.e. +11 days from the outburst, with 
 a $T$ = 7800 K blackbody]{wag04}. 
 
In quiescence or in pre-outburst phases (e.g. 2008 October 17, 2009 January 22, April 14, October 26) 
the SED indicates somewhat cooler temperatures, and a clear excess in the R-band flux can be noted. 
At these phases, the spectral lines are broader and in pure emission, He II $\lambda$4686 becomes strong
and H$\alpha$ contributes significantly to the R-band flux.

The SED of an LBV is expected to shift from the UV to the optical during outbursts. This appears to be in contradiction with
what we observe for the variable in NGC 3432 which, in outburst, moves its SED to bluer wavelengths (i.e. higher temperatures).  
On the other hand, during quiescence, the variable in NGC 3432 shows moderate temperatures \citep[$V-I \sim$ 1 would imply $T_{eff} \approx$ 5300 K,][]{dri00} 
that are typical of yellow (G-type) supergiants, and are inconsistent with early spectral  
types that LBVs are expected to show during quiescence \citep[e.g.][]{wol97}. 

The large U-band emission displayed by the variable in NGC 3432 during and soon after the eruptive events might provide key information.
Lacking UV observations, we can only speculate on  
the nature of the U-band excess. A plausible scenario is that the observed SED is in reality due to  
two variable contributions: a very hot component which peaks in the UV domain associated with the outburst, plus a warm 
component that peaks in the optical and declines in temperature with time after the outburst, 
possibly due to the hypergiant star.
A detailed study on the UV variability of the source appears to be crucial to understand the nature of the luminous star in NGC 3432.
Spectral modelling of the continuum and the strong stellar outflow would also give estimates for $T_{eff}$ and $log L$ of the star during its 
transitional phases, to determine if the observables are consistent with our ideas of LBVs expanding and cooling during outbursts.

Nevertheless, a further stellar component is still necessary to account for the presence of the He II  $\lambda$ 4686 line in the spectra 
of the variable and the high-velocity wind observed during quiescence. These observables would be more consistent with
a scenario involving also a hot Wolf-Rayet star (see Sect. \ref{prog}).

\section{The nature of the variable in NGC 3432} \label{prog} 
 
The luminous variable in NGC 3432 is one of the most intriguing stars discovered in the Local Universe. It experienced at least 4 luminous 
outbursts within a decade, but a number of minor rebrightenings were also observed (e.g. Figure \ref{fig6}). 
Since the 3 most recent outbursts occurred with intervals of about 200-210 days, one may suggest that 
subsequent eruptive events occur with some periodicity, although the modulation of the light curve is not regular enough to support this claim. 
In quiescence the  
object has an unusually high intrinsic luminosity ($M_R$ $\approx$ -10.8), consistent with that of a luminous LBV. Also, to our knowledge, its spectra are unprecedented.
Prominent H Balmer lines, dominating over other spectral lines (e.g. He I, Fe II), are typical of LBVs. However, the velocity of the ejected material, 1500-2500 km s$^{-1}$,  
is high (by a factor $\sim$ 5-10) for erupting LBVs, and are more consistent with velocities expected in winds of Wolf-Rayet stars \citep[or in some 
very hot stars, see][that are not expected to produce prominent H lines like those observed in the spectra of the NGC 3432 variable]{cas87}.  
In addition, the presence of He II during the quiescent phase is indicative of high gas temperatures, which is unusual in LBVs, and is  
commonly observed in Wolf-Rayet stars 
\cite[including Ofpe/WN9 transition types, see][]{nota96,mor96}. However, in all these stars the He II lines (in particular that at 4686\AA) are expected to dominate over all 
the other lines. This is not the case for the variable in NGC 3432, in which He II $\lambda$4686 is always less prominent than H$\beta$.
A relatively weak He II $\lambda$4686~line was occasionally observed in quiescent LBVs \citep{sta86}, but always at much lower velocities than those observed in the spectra of the variable in NGC 3432.
 
In addition, the asymmetric profile (double-peaked, see insert in Figure \ref{fig8}) of the He II $\lambda$4686 line is puzzling. It is currently unclear if this peculiar profile is due to a blend with other 
spectral lines or if it is an intrinsic asymmetry of the He II line. A similar profile in the He II $\lambda$4686 feature was observed also in 
the spectra of the luminous OB supergiant S18 in the Small Magellanic Cloud (SMC), classified as a B[e] type \citep{nota96}. In that case, the line asymmetry and its variability  
with time were interpreted as a signature of accretion onto a hot companion inside a complex, two-component circumstellar wind \citep[a hot, high-velocity 
polar wind plus a cooler, dense equatorial disk, see][]{shor87,zick89}. 
However, S18 has a moderate photometric variability\footnote{The variability of S18 is $<$1 mag, according to \citet{vange02}, 
while most of B[e]-type stars do not show significant magnitude oscillations.}, it is slightly fainter \citep[$M_{bol} \approx$ -9.2 to 
-9.4,][]{zick89} and the spectral lines are narrower than those observed in the spectra of the variable in NGC 3432. Nevertheless,  
despite the evident differences between the variable in NGC 3432 and S18 in SMC, a  
scenario involving a companion star and a complex circumstellar medium (CSM) might explain both  
the quasi-modulated variability in the light curves and the variability of the He II $\lambda$4686 line. 
 
The large photometric variability of the luminous star in NGC 3432 is another characteristic that is difficult to explain. 
A variability of 4-5 mag is not unusual for an LBV outburst. However, the frequency of these episodes is  puzzling.  
According to \citet{hum94}, different types of variability, in terms of magnitude amplitudes and time scales, are observed in LBVs:  
\begin{enumerate} 
\item {\sl giant eruptions}, like that of $\eta$ Car during the 1840s,  are rare and expected to occur in intervals of hundreds to thousands of  
years. During these outbursts the star increases its magnitude by more than 2 mag
and ejects a significant amount of its envelope. Erratic individual brightenings have short life,
although the phase of intense activity may last even a few years. This phase 
of major activity is then followed by a long period of quiescence. 
\item {\sl Eruptions} are much more frequent, being usually observed over time scales of decades. In these episodes the star 
brightens by less than 2 mag, although the bolometric luminosity remains almost constant, and the eruptive phase  
may last for months to years, followed by a luminosity minimum of similar duration. AG Car experienced this kind 
of variability during the 1980s \citep{bat,sta01}. 
\item {\sl Smaller oscillations} ($\Delta m$ $\leq$ 0.5) or even {\sl micro-variations} ($\Delta m$ $\leq$ 0.1) are also frequently  
observed in LBVs on timescales from weeks to several months. 
\end{enumerate} 
 
It is evident that the variable in NGC 3432 has unique characteristics, since it does not match any of these variability scenarios. 
Although modulated light curves (e.g. Figure \ref{fig6}) are typical of LBVs during the S-Dor phase,  the fact that
 the variable in NGC 3432 experienced at least 4 major outbursts in 9 years plus a few further minor episodes in a very short period of the 
stellar evolution (around 15 yrs) is unexpected and probably cannot be connected to an S-Dor type  
of variability. \citet{smi01} noted some similarity with the erratic fluctuations of the
light curve of V12 in NGC 2403 during the period 1949-1954 \citep{tam68,hum99}, and suggested that these oscillations may be typical of LBVs 
prior to and during a giant eruption.

Another possibility is that we are observing a very massive star in the latest stages of its life, which experiences  
a major phase of instability.  Frequent ejections of H-rich shells through a pulsational pair instability mechanism,  followed by
their collisions, have been proposed to explain the properties of the over-luminous type IIn SN 2006gy \citep{woo07}. A similar scenario has
been proposed for another type IIn event, SN 1994W \citep{des08}\footnote{Data for SN 1994W were presented by \protect\cite{sol98}. The SN was  extensively modelled also 
by \protect\cite{chug04} who interpreted its observed properties to result from the SN ejecta interacting with a dense and extended CS envelope ejected in a violent event of the progenitor star just
$\sim$1.5 years before its explosion.}. In both cases, the shell-shell interaction was able to fully explain the observed  
properties of these events, even without the need to invoke an additional contribution from radioactive $^{56}$Ni \citep[e.g.][]{smi07}. 
 
As mentioned above, the spectral lines identified in 2008-OT are consistent with those of LBVs. The spectra of the variable in NGC 3432 are indeed reminiscent to 
those of LBV NGC 2363-V1 presented in \citet{pet06}. The spectral lines and their evident P-Cygni profiles are remarkably similar in these two objects,  
but the derived velocities are about one order of magnitude lower in the LBV NGC 2363-V1. LBVs in eruption may have significantly higher velocities, but usually without 
reaching the extreme values registered in the cases of the major outbursts of the variable in NGC 3432 (see Figure \ref{fig10} panel 
{\bf B} and discussion in Sect.  \ref{sp}).   
 
The recurrent mass ejections observed in the variable in NGC 3432 are then expected to produce an increasingly  
He-rich CSM, stripping the star of its massive envelope. The above scenario, as suggested by \citet{woo07}, may also explain  
the sequence of events preceding SN 2006jc and the formation of its dense, He-rich CSM \citep{nak06,pasto07a}. Remarkably, the FWHM velocity of the H and He I lines in recent spectra 
of the variable in NGC 3432 ($\sim$2300 km s$^{-1}$) is very close to that reported by \citet{pasto07a} for the He I  
circumstellar lines of SN 2006jc (about 2200 km s$^{-1}$). This wind velocity (as mentioned above) is significantly higher than that expected for eruptions of LBVs, and more consistent 
with that of a Wolf-Rayet wind. However, H is still the most prominent spectral feature. The N and He II lines which characterize spectra of young  
Wolf-Rayet stars \citep[of WN type,][see also Sect. \ref{sp}]{cro95} are relatively weak (He II $\lambda$4686) or not detected.  
 In addition, the progenitor stars of the variable in NGC 3432 and SN 2006jc were  different. In the case of SN 2006jc,  
a young Wolf-Rayet star\footnote{Alternatively, \protect\cite{tom08} 
proposed that the precursor of SN 2006jc was a more evolved WCO star.} was proposed as progenitor \citep{pasto07a,fol07},  
while the variable in NGC 3432 is significantly less evolved, since the spectra 
are still dominated by H lines. This would suggest that it is in a transitional phase  
of its evolution, possibly approaching the Wolf-Rayet stage. 
In the light of all of this, the frequent outbursts of the NGC 3432 variable star may possibly herald a similar fate  
as the precursor of SN 2006jc and an imminent SN explosion. 
 
As an alternative to the pulsational instability, these oscillations in the light curve may indicate another plausible scenario, in which the  
NGC 3432 variable is a member of a close binary stellar system \citep[like S18 in SMC,][]{zick89} in a complex circumstellar environment.  
Binarity could  enhance the LBV instability. In that case, bipolar 
structures can be expected in the ejected material. This might be verified through high-resolution spectroscopic observations 
or with  (spectro)polarimetry, both of them unfortunately still out of reach for the variable in NGC 3432.  
 
Another object  which the luminous star in NGC 3432 may 
share some similarity with is the stellar system HD 5980 in SMC \citep[see][for a review]{koe04}. HD 5980 is one of the best studied 
stellar objects, and the brightest source \citep[with integrated magnitude $M_V$ = -7.3,][]{bp91} in the SMC. However, there is substantial evidence that it is not a single exceptionally 
massive star, but is a triple system comprising of a massive erupting LBV/WR (usually labelled as {\sl star A}), an early  
Wolf-Rayet ({\sl star B}) and, possibly, an O-type companion\footnote{According to \protect\citet{foe08} we cannot exclude that the membership of {\sl star C}  
to the stellar system HD 5980 is only a mere line-of-sight coincidence.} ({\sl star C}). The three stars have, in quiescence, individual
absolute magnitudes of about $M_V \sim$ -6 \citep{foe08}.  
{\sl Star A} \citep[with a mass of about 50 M$_\odot$,][]{nie97} and {\sl star B} ($M \approx$ 28 
M$_\odot$)\footnote{We should note that \citet{foe08} revised the classical mass estimates of \citet{nie97} proposing significantly higher masses  
for both {\sl star A} and {\sl star B}, being 58-79 M$_\odot$ and 51-67 M$_\odot$, respectively.}  
constitute a rather close, eclipsing binary system with a period of about 19 days and a rather eccentric orbit ($e$ = 0.3). 
During past decades HD 5980 gradually changed its absolute magnitude and its  spectroscopic properties. This was because {\sl star A}  
entered a very active phase of variability, during which it changed its spectrum from that of a Wolf-Rayet star \citep[of WN4-type, in early 1980s,][]{nie97} to that  
of an LBV \citep[at the time of the eruption, in 1993-1994, see][and references therein]{koe96} and back to a WN type in more recent years. During the
late 1970s to the early 1980s, the spectrum of HD 
5980 showed strong, high-velocity ($v_\infty \approx$ 2000-3000 km s$^{-1}$) He II lines which are traditionally seen in WN-type Wolf-Rayet stars.  
However, at the time of the outburst of {\sl star A}, HD 5980 increased its luminosity by about 3 mag, and the spectrum started to show prominent and  
narrower lines of H and He I \citep{bat94,bar95,koe95}, that are typical of LBV eruptions. At the time of the major eruption, the minimum wind velocities measured from the spectra  
of HD 5980 were about 500 km s$^{-1}$, but they rapidly increased in the subsequent few months to about 1000-1300 km s$^{-1}$ \citep{koe98}. 
It is generally assumed that in the pre-outburst phase the spectrum of HD 5980 was dominated by the Wolf-Rayet features attributed to {\sl star B}, 
while during outburst the spectrum was dominated by the lower-velocity LBV-like features of {\sl star A}.  However, it is still puzzling how 
{\sl star A} could be a hot WN star before (and after) its eruption as an LBV. The pieces of evidence {\bf 1)} that the spectra of HD 5980 show strong H lines (with line velocities that 
are, however, too high for an LBV) and {\bf 2)} that only one eruption has been registered so far, suggest that {\sl star A} is  not a Wolf-Rayet star in  
the classical sense, but it is in a sort of pre-LBV evolutionary phase \citep{foe08}. It is interesting to note that also a few other luminous, H-rich Wolf-Rayet  
stars are observed in the SMC \citep{foe03}.  
 
In this context, the observables of the variable star in NGC 3432 appear rather similar to those of HD 5980 during the eruptive phase. This might indicate 
that the variable in NGC 3432 is in a similar evolutionary path as {\sl star A}, i.e. in an initial LBV stage. 
This might explain the high wind velocities in an otherwise LBV-like spectrum. 
But it does not explain the strength and the variability of the He II line. In addition, 
one can speculate that some of the photometric and spectroscopic variability observed in the luminous variable in NGC 3432 can in some way be related to the close interaction with a  
companion WR star, in analogy to that observed in the stellar system HD 5980. Interestingly, both NGC 3432 and SMC have quite low oxygen abundances, of
12 + lg(O/H) = 8.3 (see Section 2) and 8.0 \citep{hun07,tru07}, respectively. 
 
Long-term monitoring of the hyper-active variable in NGC 3432 is  required to  understand the configuration of the stellar system,
the reasons of its peculiar variability and if the observed properties are an indication for a forthcoming SN explosion.
 
\section*{Acknowledgments} 

\small  
This paper is based on observations made with the Italian Telescopio Nazionale Galileo 
(TNG) operated on the island of La Palma by the Fundaci\'on Galileo Galilei of 
the INAF (Istituto Nazionale di Astrofisica); the Liverpool 
Telescope operated on the island of La Palma at the Spanish Observatorio del 
Roque de los Muchachos of the Instituto de Astrofisica de Canarias; the Nordic Optical Telescope, operated
on the island of La Palma jointly by Denmark, Finland, Iceland,
Norway, and Sweden, in the Spanish Observatorio del Roque de los
Muchachos of the Instituto de Astrofisica de Canarias; the Hale Telescope, Palomar Observatory, as part of a
continuing collaboration between the California Institute of Technology, NASA/JPL, and Cornell
University; the 2.2m telescope of  the Centro Astron\'omico Hispano Alem\'an 
(CAHA) at Calar Alto, operated jointly by the Max-Planck Institut f\"ur 
Astronomie and the Instituto de Astrof\'isica de Andaluc\'ia (CSIC); the 1.82m telescope of 
INAF-Asiago Observatory.

 This paper is also based in part on data obtained from the Isaac Newton Group Archive which is maintained as 
 part of the CASU Astronomical Data Centre at the Institute of Astronomy, Cambridge; on data collected at the Kiso observatory (University of Tokyo)  
and obtained from the SMOKA, which is operated by the Astronomy Data Center, National Astronomical Observatory of Japan and 
on observations made with the NASA/ESA Hubble Space Telescope, obtained from the data
archive at the Space Telescope Science Institute, operated by the Association of Universities for Research
in Astronomy, Incorporated, under NASA contract NAS 5-26555.
This manuscript made also use of SDSS data. Funding for the SDSS and SDSS-II has been provided by the Alfred P. Sloan Foundation, the Participating Institutions, 
the National Science Foundation, the U.S. Department of Energy, the National Aeronautics and Space Administration, the Japanese Monbukagakusho, the Max Planck Society, 
and the Higher Education Funding Council for England. The SDSS Web  Site is {\it http://www.sdss.org/}.

This manuscript used information contained in the  Bright Supernova web pages  ({\it http://www.RochesterAstronomy.org/snimages},
maintained by D. Bishop), as part of the Rochester Academy of Sciences and in the AAVSO International Database. 
We are grateful to the amateur astronomers K. Itagaki, C. Deforeit, T. Tamura,  
 G. Youman, B. H\"ausler, R. Johnson  and E. Prosperi
for sharing their images of NGC 3432 with us, to P. Holmstr\"om for his help in the observations at the Sandvreten Observatory, 
and to G. Gr\"afener for useful discussions. We also thank S. Otero and K. Weis for giving us access to the comparison data of
$\eta$ Car and 2002kg, and C. Inserra and the Padova-Asiago Supernova Group to provide us an unpublished spectrum of 2009ip.

S.B. and E.K. acknowledge partial financial support from ASI contracts `COFIS'. S.T. acknowledges support by the Trans-regional Collaborative Research Centre TRR33 "The Dark Universe" of the German Research Foundation (DFG).
S.M. acknowledges support from the Academy of Finland (project 8120503).
This work, conducted as part of the award ``Understanding the lives of massive stars from birth to supernovae'' (S.J. Smartt) made
under the European Heads of Research Councils and European Science Foundation EURYI (European Young Investigator) Awards scheme (see {\it http://www.esf.org/euryi}).

 \appendix 

\onecolumn 
\appendix 
\section{Unpublished photometry  of the variable in NGC 3432}
\small 
\begin{longtable}{cccccccc} 
  \caption{Unpublished Johnson-Bessell photometry of the variable star in NGC 3432. The numbers in brackets are the errors  
  associated to the SN photometry, and include both measurement and photometric calibration uncertainties. 
  Observations marked with the symbol ``$\star$'' are unfiltered images, and have all been 
  rescaled to R-band magnitudes according to the prescriptions of \protect\cite{pasto08}, with the only exception of 
  the observation from T. Tamura (January 4, 2001) which best matches V-band magnitudes.  
  The magnitude from T. Kryachko and S. Korotkiy's observation (October 6, 2008) is from {\it http://www.astroalert.su/2008/10/09/possible-sn-in-ngc3432/}.
  The epochs and magnitudes of the first maxima of the four major outbursts are in boldface.} \label{tab1}  
\\ \hline 
Date & avg. JD  & U & B  &  V   & R  &   I & Source \\ \hline 
1994May02 & 2449474.50 & & &  & 21.09 (0.10) & & INT$^a$  \\ 
1994May02 & 2449474.51 &  & & & 21.17 (0.16) & & INT$^a$  \\ 
1995Nov29 & 2450051    &   & $>$20.90 & $>$20.63 & $>$20.12 &$>$20.34 & KISO$^b$  \\ 
1996Mar13 & 2450156     &  & $>$20.57 & $>$20.60 & $>$20.32 &$>$19.70 & KISO$^b$  \\  
1999Dec12 & 2451524.91 &  & & & $>$19.04 & & IRO$^c$$\star$ \\ 
{\bf 2000May03} & {\bf 2451667.7} & & & & {\bf 17.4 (0.1)} & & KAIT$^d$$\star$ \\
2001Jan04 & 2451914.59 & &  & 19.42 (0.17) & & & TT$\star$ \\  
  2004May15 &  2453140.83 & &  20.72 (0.03) & 20.43 (0.03) & 19.48 (0.02) & &  HALE \\ 
  2005Mar11 &  2453440.74 & &  & $>$19.54 & $>$19.03 & & GY \\ 
  2005Mar12 &  2453441.84 & &  $>$20.04 & & & & GY \\ 
  2005Mar19 &  2453449.34 & &  &  &   $>$19.27  & &   MO$\star$ \\ 
  2005Apr06 &  2453466.71 & &  &  &   $>$20.40  & &   GY$\star$ \\ 
  2005Apr11 &  2453472.35 & &  &  &   $>$19.28  & &   MO$\star$ \\ 
  2006Apr22 &  2453848.47 & &  &  &   20.35 (0.24) & & CD$\star$ \\ 
  2006Oct28 &  2454036.60 & &  &  &   $>$19.34  & &   MO$\star$ \\ 
  2006Nov15 &  2454054.69 & &  &  &   $>$19.75  & &   MO$\star$ \\ 
  2006Dec18 &  2454087.57 & &  &  &   $>$19.62  & &   MO$\star$ \\ 
  2007Jan23 &  2454123.58 & &  &  &   19.67 (0.36) & &    MO$\star$ \\   
  2007Mar17 &  2454177.36 & &  &  &   18.45 (0.15) & &    MO$\star$ \\  
  2007Nov24 &  2454428.67 & &  &  &   $>$18.98  & &  MO$\star$ \\  
  2008Jan27 &  2454493.50 & &  &  &   $>$19.61  & &  MO$\star$ \\  
  2008Feb25 &  2454522.41 & &  &  &   $>$19.79  & &  MO$\star$ \\  
  2008Mar06 &  2454532.41 & &  &  &   19.39 (0.40) & &  MO$\star$ \\ 
  2008Mar12 &  2454537.52 & & 20.10 (0.21) & 19.75 (0.12) & 18.67 (0.20) & &  RJ\\
  2008Mar12 &  2454537.75 & &  & 19.72 (0.09) &  & &  RJ$^e$ \\
  2008Oct06 &  2454746.22 & &  &  &   17.57  & &  TKC$^f$ $\star$ \\ 
  {\bf 2008Oct07} &  {\bf 2454746.62} & &  &  &   {\bf 17.52 (0.11)} & &  MO$\star$ \\ 
  2008Oct09 &  2454749.28 & &  &  &   17.71 (0.14) & &  KI$\star$ \\ 
  2008Oct17 &  2454756.67 & &  19.98 (0.10) & 19.99 (0.11) & 18.93 (0.06) & 18.89 (0.08) & CAHA \\ 
  2008Oct17 &  2454756.71 & &  20.04 (0.28) & 19.92 (0.23) &  &  & TNG \\ 
  2008Oct17 &  2454756.72 & & & & 18.94 (0.12) &  & TNG$\star$ \\ 
  2008Oct25 &  2454764.74 & &  18.82 (0.03) & 18.59 (0.02) & 18.04 (0.02) & 17.96 (0.02) & TNG \\  
  2008Nov09 &  2454779.69 & &  22.59 (0.46) & 21.92 (0.32) & & & LT\\ 
  2008Nov16 &  2454786.51 & &  &  &   $>$20.35 & & SO$\star$ \\ 
  2008Nov19 &  2454788.52 & &  & 21.52 (0.02) & & 21.05 (0.02) & HST \\ 
  2008Nov19 &  2454789.66 & &  22.03 (0.21) & 21.67 (0.10) & & & LT\\   
  2008Nov24 &  2454794.65 & &  22.56 (0.20) & 22.13 (0.12) & 21.11 (0.12) & 21.64 (0.19) & TNG \\  
  2008Nov24 &  2454794.72 & &  22.55 (0.20) & 21.90 (0.10) & & & LT \\ 
  2008Dec09 &  2454809.62 & &  19.79 (0.03) & 19.65 (0.03) & 19.32 (0.03) & 19.25 (0.06) & EK \\   
  2008Dec10 &  2454810.61 & &  19.42 (0.07) & 19.18 (0.04) & & & LT \\  
  2008Dec17 &  2454817.67 & &  19.47 (0.02) & 19.09 (0.03) & & & LT \\  
  2008Dec21 &  2454821.72 & &  & 19.34 (0.01) & 18.79 (0.01) & & NOT \\ 
  2008Dec21 &  2454821.78 & 19.22 (0.02) &  19.77 (0.02) & & &  18.78 (0.02) & NOT \\ 
  2008Dec22 &  2454822.58 & &  19.79 (0.03) & 19.43 (0.02) & 18.84 (0.04) & 18.67 (0.02) & TNG \\  
  2008Dec23 &  2454823.50 & &  & & 18.86 (0.06) & & SO$\star$ \\ 
  2008Dec23 &  2454823.54 & &  19.80 (0.23) & 19.44 (0.13) & 18.87 (0.12) & 18.69 (0.13) & SO \\ 
  2008Dec29 &  2454829.64 & &  20.39 (0.07) & 19.85 (0.04) & & & LT \\  
  2008Dec30 &  2454830.68 & &  &  19.86 (0.07) & 19.12 (0.04) & & CAHA \\     
  2009Jan01 &  2454833.46 & &  &  & 19.22 (0.17) & & SO$\star$ \\   
  2009Jan02 &  2454833.53 & &  $>$20.24 & 20.17 (0.17) & 19.21 (0.09) & & SO \\  
  2009Jan05 &  2454836.69 & &  20.51 (0.05) & 20.05 (0.04) & & & LT \\  
  2009Jan11 &  2454842.59 & &  & & 19.09 (0.30) & & MO$\star$ \\ 
  2009Jan13 &  2454844.63 & &  20.83 (0.21) & 20.37 (0.11) & & & LT \\ 
  2009Jan14 &  2454845.62 & &  20.90 (0.07) & 20.44 (0.05) & & & LT \\   \hline
\\ 
\caption{continued.}\\ 
\hline
Date & JD  & U & B  &  V   & R  &   I & Source \\ \hline 
  2009Jan20 &  2454851.57 & &  & & 19.53 (0.06) & & TNG$\star$ \\ 
  2009Jan20 &  2454852.49 & &  21.04 (0.07) & 20.63 (0.05) & & & LT \\  
  2009Jan21 &  2454852.56 & &  21.11 (0.05) & 20.65 (0.02) & & & LT \\  
  2009Jan22 &  2454853.67 & $>$20.28 &  21.07 (0.19) & 20.83 (0.05) & 19.51 (0.02) & 19.82 (0.03) & TNG \\ 
  2009Jan25 & 2454856.55 & &  21.11 (0.05) & 20.69 (0.04) & & & LT \\ 
  2009Feb01 &  2454863.57 & $>$20.98 &  & &  & & NOT \\ 
  2009Feb01 &  2454863.57 & &  & & 19.49 (0.13) & & NOT$\star$ \\ 
  2009Feb02 &  2454865.40 & &  &  & 19.51 (0.11) & & SO$\star$ \\  
  2009Feb02 &  2454865.45 & &  &  & 19.49 (0.10) & & SO \\  
  2009Feb02 &  2454865.48 & &  21.05 (0.05) & 20.85 (0.05) & 19.48 (0.02) & 19.99 (0.03) & NOT \\ 
  2009Feb12 &  2454874.61 & &  & 20.71 (0.03) & 19.45 (0.02) & 19.81 (0.07) & NOT \\  
  2009Feb22 &  2454884.51 & &  &  & 19.56 (0.09) & & NOT \\ 
  2009Feb22 &  2454884.52 & &  &  & 19.55 (0.13) & & NOT$^g$$\star$ \\  
  2009Feb22 & 2454885.07 & $>$19.84 &  $>$20.31 & $>$19.35 & & & UVOT$^h$ \\ 
  2009Feb25 &  2454888.43 & &  21.11 (0.28) & $>$19.69 & $>$18.73 & $>$18.55 & EK$^g$ \\ 
  2009Mar15 & 2454906.46 & &  & & 19.91 (0.12) & &  SO$\star$ \\  
  2009Mar15 & 2454906.49 & &  & 20.97 (0.24) & & &  SO \\  
  2009Mar19 & 2454910.37 &  & & 21.00 (0.04) & 19.98 (0.03) & 20.15 (0.10) & CAHA\\ 
  2009Mar21 & 2454912.46 & &  & & 19.95 (0.13) & & TNG \\ 
  2009Apr14 & 2454935.62 & &  20.82 (0.04) & 20.59 (0.03) & 19.73 (0.03) & 20.03 (0.02) & NOT$\star$ \\ 
  2009Apr15 & 2454937.49 & &   & 20.06 (0.01) & 19.45 (0.01) & 19.67 (0.02) & NOT \\ 
  2009Apr22  & 2454943.58 & 17.58 (0.02) &  18.58 (0.01) & 18.45 (0.01) & 18.04 (0.01) & 18.00 (0.02) & NOT \\ 
  {\bf 2009Apr22}  & {\bf 2454944.41} & & & {\bf 18.33 (0.05)} & {\bf 17.92 (0.04)} & & SO \\ 
  2009Apr24  & 2454945.55 & & 18.60 (0.01) & 18.50 (0.02) &  18.05 (0.01) & 18.01 (0.02) & TNG \\  
  2009Apr24  & 2454946.44 & & & & 18.11 (0.11) & & BH$\star$ \\ 
  2009Apr25  & 2454947.40 & & 18.94 (0.07) & 18.81 (0.06) &  18.40 (0.05) & 18.29 (0.05) & SO \\ 
  2009Apr29  & 2454951.43 & 19.24 (0.08) & 19.85 (0.03) & 19.67 (0.02) & 19.05 (0.03) & 19.13 (0.05) & CAHA\\  
  2009May03  & 2454955.39 & & & 19.66 (0.03) & 19.07 (0.03) & 19.20 (0.06) & CAHA \\ 
  2009May05  & 2454956.55 & & & &  19.11 (0.03) & & NOT\\  
  2009May08  & 2454959.67 & & &  18.54 (0.02) & 18.05 (0.02) & 18.13 (0.02) & P60 \\ 
  2009May13  & 2454964.51 & 18.21 (0.03) & 19.00 (0.01) & 18.81 (0.01) & 18.41 (0.02) & 18.42 (0.02) & TNG \\   
  2009May19  & 2454971.38 & 19.77 (0.05) & 20.25 (0.03) & 20.07 (0.03) & 19.51 (0.03) & 19.68 (0.04) & TNG \\ 
  2009May22  & 2454974.42 & & & & $>$18.73 & & EP \\ 
  2009May25  & 2454977.36 & & 19.91 (0.07) & 19.65 (0.05) & 19.18 (0.04) & 19.05 (0.06) & EK \\ 
  2009May28  & 2454980.35 & & & & 18.97 (0.06) & & EK \\
  2009Jul16  & 2455029.33 & & & & $>$18.36 & & EP \\
  2009Jul18  & 2455031.36 & & & $>$21.01 & $>$20.98 & & CAHA \\
  2009Jul22  & 2455035.35 & & & & $>$20.58 & & EK \\
  2009Oct19  & 2455122.69 & & & 21.17 (0.38) & 20.39 (0.12) & & EK \\
  2009Oct26  & 2455130.76 & & 21.61 (0.16) & 21.36 (0.07) & 20.18 (0.04) & 20.55 (0.06) & NOT \\
  2009Nov15  & 2455150.74 & & & & 20.00 (0.17) & & NOT \\
  2009Nov19  & 2455154.67 & & & 19.66 (0.03) & 19.23 (0.02) & & CAHA \\
  2009Nov20  & 2455155.74 & & & & 18.50 (0.02) & & NOT \\
  2009Nov21  & 2455156.62 & & 19.48 (0.07) & 19.29 (0.08) & 18.73 (0.05) & 18.71 (0.08) & CAHA \\
  2009Nov22  & 2455157.76 & & 21.26 (0.12) & 20.75 (0.11) & & & LT \\
  2009Nov27  & 2455162.61 & & 19.88 (0.07) & 19.65 (0.07) & 19.23 (0.03) & 19.18 (0.03) & CAHA \\
  {\bf 2009Nov28}  & {\bf 2455163.71} & & {\bf 18.72 (0.03)} & {\bf 18.53 (0.02)} & & & LT \\ 
  2009Nov29  & 2455164.72 & & 18.83 (0.04) & 18.57 (0.03) & & & LT \\ 
  2009Nov30  & 2455165.68 & & 19.10 (0.03) & 18.93 (0.02) & & & LT \\ 
  2009Dec02  & 2455167.76 & & 19.33 (0.04) & 19.27 (0.12) & & & LT \\
  2010Jan19  & 2455215.61 & & & 20.93 (0.13) & 20.11 (0.15) & 19.75 (0.17) & EK \\
  2010Feb07 &  2455235.38 & &  &  & $>$20.30 & & SO$\star$ \\   
  2010Feb07 &  2455235.45 & &  &  & $>$20.00 & & MO$\star$ \\  
  2010Feb24  & 2455251.50 & & 21.71 (0.23) & 21.51 (0.22) & & & LT \\ 
  2010Feb24  & 2455252.47 & & 21.67 (0.31) & 21.49 (0.36) & & & LT \\
  2010Mar03  & 2455259.53 & & 22.42 (0.37) & 21.98 (0.32) & & & LT \\
  2010Mar19  & 2455274.69 & & & &  21.22 (0.19) & & NOT \\
  2010Mar21  & 2455277.41 & & 22.03 (0.21) & 21.92 (0.26) & & & LT$^i$ \\  \hline
\end{longtable} 
\hspace{-0.7cm} $^a$ ING archive: {\it http://casu.ast.cam.ac.uk/casuadc/archives/ingarch}.\\ 
$^b$ Subaru-Mitaka-Okayama-Kiso Archive System \protect\citep{baba02}.\\ 
$^c$ Template image from the Supernova Search in Late-type Galaxies using the 0.37-m Rigel Robotic Telescope  
(Winer Observatory - University of Iowa; 
{\it http://phobos.physics.uiowa.edu/images/snsearch.html}).\\ 
$^d$ Measurement from \protect\cite{wag04}. \\
$^e$ Astrodon Generation 1 Series I Luminance filter.\\
$^f$ Kryachko and Korotkiy estimated $R$ = 17.65 for 2008-OT 
with reference to star USNO-A2.0 1200-06602541 (USNO $R$ = 14.0), which is coincident with  
star D in the sequence of \protect\citet{wag04}; $R$ = 13.924. The 2008-OT magnitude was rescaled 
accordingly.\\ 
$^g$ Poor-quality night (clouds and poor seeing).\\ 
$^h$ Additional UVOT detection limits: $UVW1 >$ 19.85, $UVW2 >$ 20.18, $UVM2 >$ 19.44.\\ 
$^i$ Additional NIR detection limits were obntained with LT + SupIRCam: $J >$ 18.07 and $H >$ 17.03.\\
\\ 
INT = 2.54-m Isaac Newton Telescope + CCD-EEV5 (La Palma, Canary Islands, Spain);\\ 
KISO =  1.05-m Schmidt Telescope + TC215-1K CCD (Kiso Observatory, Inst. of Astronomy, University of Tokyo, Japan);\\ 
IRO = 0.37-m f/14 Rigel Telescope + FLI with SITe-003 CCD 
(Obs. Chris Anson, Winer Observatory, Sonoita, Arizona, US);\\ 
KAIT = 0.76-m Katzman Automatic Imaging Telescope + CCD (Lick Observatory, Mt. Hamilton, California, US); \\
TT = 0.4-m Meade LX200 Telescope + SBIG ST-8 CCD (Obs. T. Tamura, Kagawa, Japan);\\ 
HALE = Hale 5.08-m Telescope + COSMIC direct imaging 2048x2048 SITe thinned CCD (Palomar Observatory, San Diego County, California, US);\\
GY = Takahashi FS-128G Telescope + SBIG ST-10 Dual CCD Camera (Obs. G. Youman, Penryn, California, US);\\ 
MO = 0.32-m f/3.1 Newton Telescope + Starlight Xpress MX716 CCD (Obs. G.D., Moonbase Observatory, Akersberga, Sweden);\\ 
CD = 0.46-m f/2.8 Centurion Telescope + ST8E CCD camera (Obs. C. Deforeit, Caen, Normandie, France);\\ 
RJ = 0.356-m f/10 LX200R Meade Telescope + SBIG STL-11000XM CCD camera (Obs. R. Johnson, Mantrap Lake, Minnesota, US);\\
TKC = 0.3-m Ritchey-Chr\'etien Telescope + Apogee Alta U9000 CCD (Kazan State Univ., Astrotel Obs., Karachay-Cherkessia, Russia);\\ 
KI = 0.60-m f/5.7 Reflector Telescope + KAF-1001E 1024$\times$1024pixel CCD (Obs. Koichi Itagaki, Yamagata, Japan);\\ 
CAHA =  2.2-m Calar Alto Telescope + CAFOS (German-Spanish Astron. Center, Sierra de Los Filabres, Andaluc\'{i}a, Spain);\\ 
TNG = 3.58-m Telescopio Nazionale Galileo + Dolores (Fundaci\'{o}n Galileo Galilei-INAF, La Palma, Canary Islands, Spain);\\ 
LT = 2.0-m Liverpool Telescope + RatCAM (La Palma, Canary Islands, Spain);\\ 
SO = 0.44-m f/4.4 Newton Telescope + SBIG ST-7E Dual CCD (Obs. L.H., Sandvreten Obs., Uppsala, Sweden);\\ 
HST = Hubble Space Telescope + WFPC camera (P.I. W. Li; proposal ID:10877); \\ 
EK = 1.82-m Copernico Telescope + AFOSC (INAF - Osservatorio Astronomico di Asiago, Mt. Ekar, Asiago, Italy);\\ 
NOT = 2.56m Nordic Optical Telescope + ALFOSC (La Palma, Canary Islands, Spain);\\
UVOT = Swift Gamma-Ray Burst Explorer + UVOT (proposal ID 00036566001); \\ 
BH = Meade LX200 12" f-6.3 telescope + SBIG ST-10 Dual CCD Camera (Obs. B. H\"ausler, Rimpar, Germany);\\ 
P60 = Palomar 60-inch Telescope + CCD camera (Palomar Mountain,  San Diego County, California, US);\\ 
EP = Meade LX200 12" f-6.3 telescope + ST10XME CCD (Obs. E. Prosperi, Osservatorio Remoto Skylive,  Pedara, Catania, Italy). 
\twocolumn 


\begin{table*} 
 \centering 
 \begin{minipage}{140mm} 
  \caption{Sloan-filter photometry of the variable star in NGC 3432 and associated errors.} \label{tab2} 
  \begin{tabular}{ccccccccc} 
\\  \hline 
Date & JD  &  u  &  g   & r  &   i & z & Source \\ \hline 
  2003Apr26 & 2452755.72 & 20.39 (0.09) & 19.99 (0.04) & 19.20 (0.03) & 19.57 (0.08) & 19.54 (0.10) & SDSS \\  
  2004Mar16 & 2453080.89 & $>$22.37     & 22.26 (0.29) & 21.46 (0.13) & 21.51 (0.17) & 21.23 (0.37) & SDSS \\ 
  2008Nov09 & 2454779.69 & & 21.87 (0.32) & 20.67 (0.14) & $>$21.37 & $>$20.90 & LT \\  
  2008Nov19 & 2454789.66 & & 22.02 (0.24) & 20.90 (0.10) & 21.54 (0.19) & $>$21.89 & LT \\  
  2008Nov24 & 2454794.66 & & 22.12 (0.20) & 21.31 (0.11) & 21.52 (0.10) &  & LT \\  
  2008Dec10 & 2454810.61 & & & 18.79 (0.06) & 19.17 (0.05) & & LT \\  
  2008Dec17 & 2454817.67 & & & 18.79 (0.02) & 18.94 (0.03) & & LT \\ 
  2008Dec22 & 2454822.55 & & & 18.99 (0.03) & & & LT \\ 
  2008Dec29 & 2454829.64 & & & 19.23 (0.02) & 19.53 (0.03) & & LT \\ 
  2009Jan05 & 2454836.60 & & & 19.31 (0.02) & & & LT \\  
  2009Jan05 & 2454836.69 & & & 19.30 (0.03) & 19.63 (0.03) & & LT \\  
  2009Jan13 & 2454844.63 & & & 19.43 (0.03) & 19.94 (0.12) & & LT \\ 
  2008Jan14 & 2454845.62 & & & 19.60 (0.03) & 20.04 (0.03) & & LT\\  
  2008Jan20 & 2454851.54 & & & 19.58 (0.02) & & & LT \\ 
  2008Jan20 & 2454852.49 & & & 19.53 (0.05) & 20.12 (0.06) & & LT \\ 
  2008Jan21 & 2454852.56 & & & 19.57 (0.02) & 20.18 (0.05) & & LT \\ 
  2009Jan25 & 2454856.55 & & & 19.61 (0.02) & 20.24 (0.04) & & LT \\ 
  2009Jan29 & 2454860.51 & & & 19.62 (0.02) & & & LT \\ 
  2009Feb16 & 2454879.40 & & & 19.66 (0.06) & & & LT \\ 
  2009Feb16 & 2454884.44 & & & 19.71 (0.11) & & & LT \\ 
  2009Mar09 & 2454900.41 & & & 19.85 (0.09) & & & LT \\  
  2009Mar23 & 2454914.38 & & & 20.08 (0.02) & & & LT \\  
  2009Mar30 & 2454921.36 & & & 20.11 (0.03) & & & LT \\  
  2009Apr06 & 2454928.36 & & & 20.13 (0.07) & & & LT \\  
  2009Apr13 & 2454935.37 & & & 19.90 (0.03) & & & LT \\ 
  2009Apr21 & 2454943.35 & & & 18.24 (0.04) & & & LT \\ 
  2009Apr23 & 2454945.42 & & & 18.29 (0.01) & & & LT \\ 
  2009Apr28 & 2454950.46 & & & 19.05 (0.01) & & & LT \\ 
  2009Apr29 & 2454951.48 & & & 19.15 (0.02) & & & LT \\ 
  2009May06 & 2454957.53 & & & 18.33 (0.02) & & & LT \\ 
  2009May11 & 2454963.39 & & & 19.99 (0.02) & & & LT \\ 
  2009May14 & 2454966.39 & & & 19.79 (0.02) & & & LT \\ 
  2009May17 & 2454969.39 & & & 20.30 (0.04) & & & LT \\ 
  2009May20 & 2454972.38 & & & 19.37 (0.04) & & & LT \\ 
  2009May23 & 2454975.39 & & & 19.20 (0.02) & & & LT \\ 
  2009May26 & 2454978.39 & & & 19.35 (0.03) & & & LT \\  
  2009May29 & 2454981.40 & & & 18.97 (0.03) & & & LT \\  
  2009Jun09 & 2454992.43 & & & 18.95 (0.02) & & & LT \\  
  2009Jun12 & 2454995.39 & & & 19.73 (0.04) & & & LT \\  
  2009Jun14 & 2454997.39 & & & 20.37 (0.08) & & & LT \\  
  2009Jun22 & 2455005.39 & & & 20.99 (0.05) & & & LT \\   
  2009Jun23 & 2455006.43 & & & 21.11 (0.12) & & & LT \\   
  2009Jun29 & 2455012.40 & & & 21.35 (0.10) & & & LT \\   
  2009Jul08 & 2455021.39 & & & 21.31 (0.17) & & & LT \\   
  2009Jul09 & 2455022.38 & & &   $>$21.18   & & & LT \\   
  2009Jul12 & 2455025.40 & & & 21.68 (0.12) & & & LT \\   
  2009Jul17 & 2455030.38 & & &   $>$21.37   & & & LT \\ 
  2009Jul18 & 2455031.38 & & &   $>$21.27   & & & LT \\ 
  2009Nov14 & 2455149.75 & & & 20.34 (0.04) & & & LT \\ 
  2009Nov22 & 2455157.75 & & & 19.96 (0.12) & & & LT \\
  2009Nov22 & 2455157.76 & 21.42 (0.19) & 20.87 (0.12) & 20.01 (0.08) & 20.52 (0.18) & 20.41 (0.34) & LT \\
  2009Nov24 & 2455159.75 & & & 20.82 (0.10) & & & LT \\ 
  2009Nov25 & 2455160.74 & & & 20.62 (0.06) & & & LT \\ 
  2009Nov26 & 2455161.67 & & & 20.70 (0.12) & & & LT \\ 
  2009Nov27 & 2455162.75 & & & 19.03 (0.02) & & & LT \\ 
  2009Nov28 & 2455163.71 & 18.42 (0.06) & 18.61 (0.08) & 18.32 (0.02) & 18.48 (0.04) & 18.59 (0.04) & LT \\ 
  2009Nov29 & 2455164.72 & 18.69 (0.03) & 18.71 (0.09) & 18.35 (0.03) & 18.52 (0.04) & 18.54 (0.04) & LT \\ 
  2009Nov30 & 2455165.68 & 18.88 (0.05) & 18.96 (0.03) & 18.61 (0.03) & 18.90 (0.03) & 18.81 (0.05) & LT \\
  2009Dec02 & 2455167.76 & 19.53 (0.09) & 19.28 (0.06) & 18.79 (0.05) & 18.96 (0.07) & 19.02 (0.13) & LT \\
  2009Dec09 & 2455174.61 & & & 19.93 (0.06) & & & LT \\ 
  2009Dec27 & 2455192.56 & & & 20.41 (0.24) & & & LT \\ 
  2010Feb24 & 2455251.50 & $>$21.75 & &  20.96 (0.10) & 20.97 (0.17) & 21.09 (0.22) & LT \\
  2010Feb24 & 2455252.47 & $>$21.94 & &  21.14 (0.22) & 21.03 (0.16) & 21.28 (0.33) & LT \\
  2010Mar04 & 2455259.53 & $>$22.23 & &  21.59 (0.35) & 21.55 (0.37) & 21.49 (0.46) & LT \\  
  2010Mar21 & 2455277.41 & $>$22.03 & &  21.36 (0.13) & 21.32 (0.32) & $>$20.82 & LT \\  
  \hline 
 \end{tabular} 
\hspace{-0.7cm} 
 
 SDSS = 2.5-m Sloan Digital Sky Survey Telescope + TK2048E CCDs (Apache Point Observatory, New Mexico, USA);\\ 
 LT = 2.0-m Liverpool Telescope + RatCAM (La Palma, Canary Islands, Spain). 
\end{minipage} 
\end{table*} 

 \normalsize 
 
 
 

\normalsize

\begin{thebibliography}{99} 
\bibitem[\protect\citeauthoryear{Agnoletto et al.}{2009}]{agn08} Agnoletto, I. et al., 2009, ApJ, 691, 1348 
\bibitem[\protect\citeauthoryear{Aretxaga et al.}{1999}]{are99} Aretxaga, I., Benetti, S., Terlevich, R. J., Fabian, A. C., Cappellaro, E., Turatto, M.,  
della Valle, M. 1999, MNRAS, 309, 343 
\bibitem[\protect\citeauthoryear{Arp}{1966}]{arp66} Arp, H. C. 1966, {\it Atlas of Peculiar Galaxies} (Caltech: Pasadena)  
\bibitem[\protect\citeauthoryear{Bateson}{1988-2000}]{bat} Bateson, F. M. 1988-2000, RASNZ Var. Star Sect. Circ. 
\bibitem[\protect\citeauthoryear{Bateson et al.}{1994}]{bat94} Bateson, F. M., Gilmore, A., Jones, A. 1994,  IAU Circ. 6102 
\bibitem[\protect\citeauthoryear{Baba et al.}{2002}]{baba02} Baba, H. et al. 2002, ADASS XI, eds. D. A. Bohlender, D. Durand, \& T. H. Handley, ASP Conference
Series, Vol.281, p.298 
\bibitem[\protect\citeauthoryear{Barb\'a et al.}{1995}]{bar95} Barb\'a R. H., Niemela, V. S., Baume, G., V\'asquez, R. A., 1995, ApJ, 446, L23 
\bibitem[\protect\citeauthoryear{Berger et al.}{2009}]{ber09} Berger, E. et al. 2009, ApJ, 699, 1850
\bibitem[\protect\citeauthoryear{Blondin et al.}{2006}]{blo06} Blondin, S., Masters, K., Modjaz, M., Kirshner, R., Challis, P., 
Matheson, T., Berlind, P. 2006, CBET, 636, 1 
\bibitem[\protect\citeauthoryear{Bond et al.}{2009}]{bond09} Bond, H. E., Bedin, L. R., Bonanos, A. Z., Humphreys, R. M., Monard, B. L. A. G.; Prieto, J. L., Walter, F. M. 2009, ApJ, 695, L154 
\bibitem[\protect\citeauthoryear{Botticella et al.}{2009}]{bot08} Botticella, M. T. et al . 2009, MNRAS, 398, 1041
\bibitem[\protect\citeauthoryear{Branch \& Cowan}{1985}]{bra85} Branch, D., Cowan, J. J. 1985, ApJ, 297L, 33 
\bibitem[\protect\citeauthoryear{Breysacher \& Perrier}{1991}]{bp91}   Breysacher, J., Perrier, C. 1991, in IAU Symp. 143,  
{\it Wolf-Rayet Stars and Interrelations with Other Massive Stars in Galaxies}, ed. K. van der Hucht \& B. Hidayat, Kluwer Academic Publishers, Dordrecht, p.229 
\bibitem[\protect\citeauthoryear{Cassinelli \& Lamers}{1987}]{cas87} Cassinelli, J. P., Lamers, H. J. G. L. M. 1987, in {\it 
Exploration of the Universe with the IUE Satellite}, ed. Y. Kondo, (Dordrecht: Reidel), p.139 
\bibitem[\protect\citeauthoryear{Chu et al.}{2004}]{chu04} Chu, Y.-H., Gruendl, R. A., Stockdale, C. J., Rupen, M. P., Cowan, J. J.,  
Teare, S. W. 2004, AJ, 127, 2850 
\bibitem[\protect\citeauthoryear{Chugai et al.}{2004}]{chug04} Chugai, N. N. et al. 2004, MNRAS, 352, 1213
\bibitem[\protect\citeauthoryear{Clark et al.}{2009}]{cla09} Clark, J. S., Crowther, P. A., Mikles, V. J. 2009, A\&A 507, 1567
\bibitem[\protect\citeauthoryear{Cowan et al.}{1988}]{cow88} Cowan, J. J., Henry, R. B. C., Branch, D., 1988, ApJ, 329, 116 
\bibitem[\protect\citeauthoryear{Crowther et al.}{1995a}]{cro95} Crowther, P. A., Smith, I. J., Hillier, D. J., Schmutz, W. 1995a, A\&A, 293, 427 
\bibitem[\protect\citeauthoryear{Crowther et al.}{1995b}]{cro95b} Crowther, P. A., Hillier, D. J., Smith, L. J. 1995b, A\&A, 293, 403 
\bibitem[\protect\citeauthoryear{Crowther et al.}{1995c}]{cro95c} Crowther, P. A., Hillier, D. J., Smith, L. J. 1995c, A\&A, 293, 172 
\bibitem[\protect\citeauthoryear{Davidson et al.}{1995}]{dav95} Davidson, K., Ebbets, D., Weigelt, G., Humphreys, R. M., Hajian, A. R.,  
Walborn, N. R., Rosa, M.  1995, AJ, 109, 1784 
\bibitem[\protect\citeauthoryear{Davidson \&  Humphreys}{1997}]{dav97} Davidson, K., Humphreys, R. M. 1997, AAR\&A, 35, 1  
\bibitem[\protect\citeauthoryear{Dessart et al.}{2009}]{des08} Dessart, L., Hillier, D. J., Gezari, S., Basa, S., Matheson, T. 2009, 
MNRAS, 349, 21 
\bibitem[\protect\citeauthoryear{Drilling \& Landolt}{2000}]{dri00}   Drilling, J. S., Landolt, A. U. 2000, in
{\it Allen's astrophysical quantities}, 4th ed. Edited by Arthur N. Cox. ISBN: 0-387-98746-0. Publisher: New York: AIP Press; Springe r, 2000, p.381
\bibitem[\protect\citeauthoryear{Duszanowicz et al.}{2008}]{dus08} Duszanowicz, G., Nakano, S., Itagaki, K. 2008, CBET, 1534, 1 
\bibitem[\protect\citeauthoryear{Elmhamdi et al.}{2003}]{elm03} Elmhamdi, A. et al. 2003, MNRAS, 338, 939 
\bibitem[\protect\citeauthoryear{English \& Irwin}{1997}]{eng97} English, J., Irwin, J. A. 1997, AJ, 113, 2006 
\bibitem[\protect\citeauthoryear{Evans et al.}{2006}]{evan06} Evans, C. J., Lennon, D. J., Smartt, S. J., Trundle, C. 2006, A\&A, 
456, 623 
\bibitem[\protect\citeauthoryear{Fesen}{1985}]{fes85} Fesen, R. A. 1985, ApJ, 297L, 29 
\bibitem[\protect\citeauthoryear{Filippenko et al.}{1995}]{fil95} Filippenko, A. V., Barth, A. J., Bower, G. C., Ho, L. C.,  
Stringfellow, G. S., Goodrich, R. W., Porter, A. C. 1995, AJ, 110, 2261 
\bibitem[\protect\citeauthoryear{Filippenko et al.}{1999}]{fil99} Filippenko, A. V., Li, W. D., Modjaz, M. 1999, IAU Circ. 7152, 2 
\bibitem[\protect\citeauthoryear{Filippenko et al.}{2001}]{fil01} Filippenko, A. V., Li, W. D., Treffers, R. R., Modjaz, M. 2001, 
{\it Small-Telescope Astronomy on Global Scales}, ASP Conf. Ser. 246, ed. W. P. Chen, C. Lemme, B. Paczy\'nski (San Francisco: ASP), 121  
\bibitem[\protect\citeauthoryear{Foellmi et al.}{2003}]{foe03} Foellmi, C., Moffat, A. F. J., Guerrero, M. A. 2003, MNRAS, 338, 1025 
\bibitem[\protect\citeauthoryear{Foellmi et al.}{2008}]{foe08} Foellmi, C. et al. 2008, RevMexAA, 44, 3   
\bibitem[\protect\citeauthoryear{Foley et al.}{2007}]{fol07} Foley, R. J. et al., 2007, ApJ, 657, L105 
\bibitem[\protect\citeauthoryear{Foley et al.}{2010}]{fol10} Foley, R. J. et al., 2010, ApJ submitted (arXiv:1002.0635)
\bibitem[\protect\citeauthoryear{Fransson et al.}{2002}]{fra02} Fransson, C. et al. 2002, ApJ, 572, 350 
\bibitem[\protect\citeauthoryear{Freedman et al.}{2001}]{fre01} Freedman, W. L. et al. 2001, ApJ, 553, 47
\bibitem[\protect\citeauthoryear{Frew}{2004}]{fre04} Frew, D. J. 2004, JAD, 10, 6 
\bibitem[\protect\citeauthoryear{Fukugita et al.}{1996}]{fuk96} Fukugita, M., Ichikawa, T., Gunn, J. E., Doi, M., Shimasaku, K., and Schneider, D. P.  
1996, AJ, 111, 1748 
\bibitem[\protect\citeauthoryear{Gal-Yam et al.}{2007}]{gal07} Gal-Yam, Avishay, Leonard, D. C., Fox, D. B., Cenko, S. B. 
2007, ApJ, 656, 372 
\bibitem[\protect\citeauthoryear{Gal-Yam \& Leonard}{2009}]{gal08} Gal-Yam, Avishay, Leonard, D. C. 2009, Nature, 458, 865
\bibitem[\protect\citeauthoryear{Goodrich et al.}{1989}]{goo89} Goodrich, R. W., Stringfellow, G. S., Penrod, G. D., Filippenko, A. V. 
1989, ApJ, 342, 908 
\bibitem[\protect\citeauthoryear{Groh et al.}{2009}]{gro09} Groh, J. H., Hillier, D. J., Damineli, A., Whitelock, P. A., Marang, F., Rossi, C. 2009, ApJ,
698, 1698
\bibitem[\protect\citeauthoryear{Hamann et al.}{1995a}]{ham95a} Hamann, W.-R., Koesterke, L., Wessolowski, U. 
1995a, A\&A, 229, 151 
\bibitem[\protect\citeauthoryear{Hamann et al.}{1995b}]{ham95b} Hamann, W.-R., Koesterke, L., Wessolowski, U. 
1995b, A\&AS, 113, 459 
\bibitem[\protect\citeauthoryear{Harutyunyan et al.}{2007}]{har07} Harutyunyan, A., Navasardyan, H., Benetti, S., Turatto, M., 
Pastorello, A., Taubenberger, S. 2007, CBET 1184, 1 
\bibitem[\protect\citeauthoryear{Ho et al.}{1997}]{ho97} Ho, L. C., Filippenko, A. V., Sargent, W. L. W. 1997, ApJS, 112, 315 
\bibitem[\protect\citeauthoryear{Humphreys \& Davidson}{1994}]{hum94} Humphreys, R. M., Davidson, K. 1994, PASP, 106, 1025 
\bibitem[\protect\citeauthoryear{Humphreys et al.}{1999}]{hum99} Humphreys, R. M., Davidson, K., Smith, N. 1999, PASP, 111, 1124 
\bibitem[\protect\citeauthoryear{Hunter et al.}{2007}]{hun07} Hunter, I. et al. 2007, A\&A, 466, 277 
\bibitem[\protect\citeauthoryear{Koenigsberger et al.}{1995}]{koe95} Koenigsberger, G., Guinan, E., Auer, L., Georgiev, L. 1995, ApJ, 452, 107 
\bibitem[\protect\citeauthoryear{Koenigsberger et al.}{1996}]{koe96} Koenigsberger, G., Guinan, E., Auer, L., Georgiev, L. 1996, RevMexAA, 5, 92 
\bibitem[\protect\citeauthoryear{Koenigsberger et al.}{1998}]{koe98} Koenigsberger, G., Pena, M., Schmutz, W., Ayala, S. 1998, ApJ, 499, 889 
\bibitem[\protect\citeauthoryear{Koenigsberger}{2004}]{koe04} Koenigsberger, G. 2004, RevMexAA, 40, 107 
\bibitem[\protect\citeauthoryear{Koenigsberger et al.}{2006}]{koe06} Koenigsberger, G., Fullerton, A. W., Massa, D., Auer, L. H. 2006, AJ, 132,
1527
\bibitem[\protect\citeauthoryear{Kotak \& Vink}{2006}]{kot06} Kotak, R., Vink, J. S. 2006, A\&A, 460L, 5 
\bibitem[\protect\citeauthoryear{Kulkarni et al.}{2007}]{kul07} Kulkarni, S. R. et al. 2007, Nature, 447, 458 
\bibitem[\protect\citeauthoryear{Lamers et al.}{1995}]{lam95} Lamers, H. J. G. L. M., Snow, T. P., Lindholm, D. M. 1995, ApJ, 455, 269
\bibitem[\protect\citeauthoryear{Lee et al.}{2009}]{lee09} Lee, J. C. et al. 2009, ApJ, 706, 599
\bibitem[\protect\citeauthoryear{Li et al.}{2002}]{li02} Li, W., Filippenko, A. V., Van Dyk, S. D., Hu, J., Qiu, Y., Modjaz, M.,  
Leonard, D. C. 2002, PASP, 114, 403L  
\bibitem[\protect\citeauthoryear{Matheson \& Calkins}{2001}]{mat01} Matheson, T., Calkins, M. 2001, IAU Circ. 7597, 3 
\bibitem[\protect\citeauthoryear{Maund et al.}{2006}]{mau06} Maund, J. R. et al., 2006, MNRAS, 369, 390 
\bibitem[\protect\citeauthoryear{Maund et al.}{2008}]{mau08} Maund, J. R. et al., 2008, MNRAS, 387, 1344 
\bibitem[\protect\citeauthoryear{Mihos \& Hernquist}{1994}]{mih94} Mihos, J. C., Hernquist, L. 1994, ApJ, 425, L13 
\bibitem[\protect\citeauthoryear{Morris et al.}{1996}]{mor96} Morris, P. W., Eenens, P. R. J., Hanson, M. M., Conti, P. S., Blum, R. D.  
1996, ApJ, 470, 597 
\bibitem[\protect\citeauthoryear{Nakano et al.}{2006}]{nak06} Nakano, S., Igataki, K., Puckett, T., Gorelli, R. 2006,  
CBET 666, 1 
\bibitem[\protect\citeauthoryear{Niemela et al.}{1997}]{nie97} Niemela, V. S., Barb\'a, R. H., Morrell, N. I., Corti, M. 
1997, in ASP Conf. Ser. 120, {\it Luminous Blue Variables: Massive Stars in Transition}, ed. A. Nota \& H. Lamers (S. Francisco: ASP), 222 
\bibitem[\protect\citeauthoryear{Nota et al.}{1996}]{nota96} Nota, A., Pasquali, A., Drissen, L., Leitherer, C., Robert, C.,  
Moffat, A. F. J., Schmutz, W. 1996, ApJS, 102, 383 
\bibitem[\protect\citeauthoryear{Ofek et al.}{2008}]{ofek08} Ofek, E. O. et al. 2008, ApJ, 674, 447 
\bibitem[\protect\citeauthoryear{Pastorello et al.}{2002}]{pasto02} Pastorello, A. et al. 2002, MNRAS, 333, 27 
\bibitem[\protect\citeauthoryear{Pastorello et al.}{2005}]{pasto05}   Pastorello, A., Aretxaga, I., Zampieri, L., Mucciarelli, P.,  
Benetti, S. 2005, in {\it 1604-2004: Supernovae as Cosmological Lighthouses}, ASP Conference Series, Vol. 342,  
Proceedings of the conference held 15-19 June, 2004 in Padua, Italy. Edited by M. Turatto, S. Benetti, L. Zampieri, and W. Shea.  
San Francisco: Astronomical Society of the Pacific, 2005., p.285 
\bibitem[\protect\citeauthoryear{Pastorello et al.}{2007a}]{pasto07a} Pastorello, A. et al. 2007a, Nature 447, 829 
\bibitem[\protect\citeauthoryear{Pastorello et al.}{2007b}]{pasto07b} Pastorello, A. et al. 2007b, Nature 449, 1 
\bibitem[\protect\citeauthoryear{Pastorello et al.}{2008}]{pasto08} Pastorello, A. et al. 2008, MNRAS, 389, 955 
\bibitem[\protect\citeauthoryear{Papenkova \& Li}{2000}]{pap00} Papenkova, M., Li W. 2000, IAU Circ. 7415, 1 
\bibitem[\protect\citeauthoryear{Petit et al.}{2006}]{pet06} Petit, V., Drissen, L., Crowther, P. A. 2006, AJ, 132, 1756 
\bibitem[\protect\citeauthoryear{Pilyugin \& Thuan}{2007}]{pyl07} Pilyugin, L. S., Thuan, T. X. 2007, ApJ, 669, 299 
\bibitem[\protect\citeauthoryear{Prieto et al.}{2008a}]{pri08a} Prieto, J. L. et al. 2008, ApJ, 681L, 9 
\bibitem[\protect\citeauthoryear{Prieto et al.}{2008b}]{pri08b} Prieto, J. L., Kistler, M. D., Stanek, K. Z., Thompson, T. A., Kochanek, C. S.,  
Beacom, J. F. 2008, Astron. Tel. 1596, 1 
\bibitem[\protect\citeauthoryear{Prinja et al.}{1990}]{pri90} Prinja, R. K., Barlow, M. J., Howarth, I. D. 1990, ApJ, 361, 607
\bibitem[\protect\citeauthoryear{Pumo et al.}{2009}]{pumo09} Pumo, M. L. et al. 2009, ApJ, 705, L138
\bibitem[\protect\citeauthoryear{Rau et al.}{2007}]{rau07} Rau, A., Kulkarni, S. R., Ofek, E. O., Yan, L. 2007, ApJ, 659, 1536 
\bibitem[\protect\citeauthoryear{Schlegel et al.}{1998}]{sch98}   Schlegel, D. J., Finkbeiner, D. P., Davis, M. 1998, ApJ, 500, 525 
\bibitem[\protect\citeauthoryear{Schinzel et al.}{2009}]{sci08} Schinzel, F. K., Taylor, G. B., Stockdale, C. J., Granot, J.,  
Ramirez-Ruiz, E. 2009, ApJ, 691, 1380
\bibitem[\protect\citeauthoryear{Shore et al.}{1987}]{shor87} Shore, S. N., Sanduleak, N., Allen, D. A. 1987, A\&A, 176, 59 
\bibitem[\protect\citeauthoryear{Smith et al.}{2002}]{smt02} Smith, J.A. et al. 2002, AJ, 123, 2121 
\bibitem[\protect\citeauthoryear{Smith et al.}{2001}]{smi01} Smith, N., Humphreys, R. M., Gehrz, R. D. 2001, PASP, 113, 692 
\bibitem[\protect\citeauthoryear{Smith et al.}{2007}]{smi07} Smith, N. et al. 2007, ApJ, 666, 1116 
\bibitem[\protect\citeauthoryear{Smith et al.}{2008}]{smi08} Smith, N., Chornock, R., Li, W., Ganeshalingam, M.,  
Silverman, J. M., Foley, R. J., Filippenko, A. V., Barth, A. J 2008, 686, 467  
\bibitem[\protect\citeauthoryear{Smith et al.}{2009}]{smi09} Smith, N. et al. 2009, ApJ, 697L, 49 
\bibitem[\protect\citeauthoryear{Smith et al.}{2010}]{smi09b} Smith, N. et al. 2010, AJ, 139, 1451 
\bibitem[\protect\citeauthoryear{Sollerman et al.}{1998}]{sol98} Sollerman, J., Cumming, R. J., Lundqvist, P. 1998, ApJ, 493, 933
\bibitem[\protect\citeauthoryear{Stahl}{1986}]{sta86} Stahl, O. 1986, A\&A, 164, 321
\bibitem[\protect\citeauthoryear{Stahl et al.}{2001}]{sta01} Stahl, O., Jankovics, I., Kov\'acs, J., Wolf, B., Schmutz, W., Kaufer, A., Rivinius, 
Th., Szeifert, Th. 2001, A\&A, 375, 54 
\bibitem[\protect\citeauthoryear{Stahl et al.}{2003}]{sta03} Stahl, O., G\"ang, T., Sterken, C., Kaufer, A., Rivinius, T., Szeifert, T., Wolf, B. 2003 
A\&A, 400, 279 
\bibitem[\protect\citeauthoryear{Stanishev}{2007}]{stan07} Stanishev, V. 2007, AN, 328, 948 
\bibitem[\protect\citeauthoryear{Stockdale et al.}{2001}]{sto01} Stockdale, C. J., Rupen, M. P., Cowan, J. J., Chu, Y.-H., Jones, S. S. 
2001, AJ, 122, 283 
\bibitem[\protect\citeauthoryear{Swaters et al.}{2002}]{swa02} Swaters, R. A., van Albada, T. S., van der Hulst, J. M., Sancisi, R. 2002, 
A\&A, 390, 829 
\bibitem[\protect\citeauthoryear{Tammann \& Sandage}{1968}]{tam68} Tammann, G. A., Sandage, A., 1968, ApJ, 151, 825 
\bibitem[\protect\citeauthoryear{Terry et al.}{2002}]{ter02}  Terry, J. N., Paturel, G., Ekholm, T. 2002, A\&A, 393, 57
\bibitem[\protect\citeauthoryear{Thompson et al.}{2009}]{tho08} Thompson, T. A., Prieto, J. L., Stanek, K. Z., Kistler, M. D., Beacom, J. F.,  
Kochanek, C. S. 2009, ApJ, 705, 1364
\bibitem[\protect\citeauthoryear{Tominaga et al.}{2008}]{tom08} Tominaga, N. et al. 2008, ApJ, 687, 1208 
\bibitem[\protect\citeauthoryear{Trundle et al.}{2007}]{tru07} Trundle, C., Dufton, P. L., Hunter, I., Evans, C. J., Lennon, D. J.,  
Smartt, S. J., Ryans, R. S. I. 2007, A\&A, 471, 625 
\bibitem[\protect\citeauthoryear{Trundle et al.}{2008}]{tru08} Trundle, C., Kotak, R., Vink, J. S., Meikle, W. P. S. 2008, A\&A, 483L, 47 
\bibitem[\protect\citeauthoryear{Turatto et al.}{1993}]{tura93} Turatto, M., Cappellaro, E., Danziger, I. J., Benetti, S., 
Gouiffes, C., della Valle, M. 1993, MNRAS, 262, 128 
\bibitem[\protect\citeauthoryear{Utrobin}{1984}]{utr84} Utrobin, V. P. 1984, Ap\&SS, 98, 115 
\bibitem[\protect\citeauthoryear{van der Kruit}{2007}]{vdk07} van der Kruit, P. C. 2007, A\&A, 466, 883 
\bibitem[\protect\citeauthoryear{van Dyk et al.}{2000}]{van00} Van Dyk, S. D., Peng, C. Y., King, J. Y., Filippenko, A. V., Treffers, 
R. R., Li, W., Richmond, M. W. 2000, PASP, 112, 1532 
\bibitem[\protect\citeauthoryear{van Dyk et al.}{2002}]{van02} Van Dyk, S. D., Filippenko, A. V., Li, W. 2002, PASP, 114, 700 
\bibitem[\protect\citeauthoryear{van Dyk et al.}{2005}]{van05} Van Dyk, S. D., Filippenko, A. V., Chornock, R., Li, W., Challis, P. M. 
2005, PASP, 117, 553 
\bibitem[\protect\citeauthoryear{van Genderen}{2001}]{vge01} van Genderen, A. M. 2001, A\&A, 366, 508 
\bibitem[\protect\citeauthoryear{van Genderen \& Sterken}{2002}]{vange02} van Genderen, A. M., Sterken, C. 2002, A\&A, 386, 926 
\bibitem[\protect\citeauthoryear{Vink et al.}{1999}]{vink99} Vink, J. S., de Koter, A., Lamers, H. J. G. L. M. 1999, A\&A, 350, 181
\bibitem[\protect\citeauthoryear{Vink \& de Koter}{2002}]{vink02} Vink, J, S., de Koter, A. 2002. A\&A, 393, 543
\bibitem[\protect\citeauthoryear{Vink et al.}{2009}]{vink09} Vink, J. S., Davies, B., Harries, T. J., Oudmaijer, R. D., Walborn, N. R. 2009,
A\&A, 505, 743
\bibitem[\protect\citeauthoryear{Wagner et al.}{2004a}]{wag04} Wagner, R. M. et al. 2004a, PASP, 116, 326 
\bibitem[\protect\citeauthoryear{Wagner et al.}{2004b}]{wag04b} Wagner, R. M., Schmidt, G. D., Smith, P., Hines, D., Starrfield, S. G.  
2004b, IAU Circ. 7417, 2 
\bibitem[\protect\citeauthoryear{Wanajo et al.}{2009}]{wan08} Wanajo, S., Nomoto, K., Janka, H.-Th., Kitaura, F. S., Mueller, B. 2009, ApJ, 695, 208 
\bibitem[\protect\citeauthoryear{Weis \& Bomans}{2005}]{wei05} Weis, K., Bomans, D. J. 2005, A\&A, 429L, 13 
\bibitem[\protect\citeauthoryear{Wolf \& Kaufer}{1997}]{wol97} Wolf, B., Kaufer, A. 1997, in {\it Luminous Blue Variables: Massive 
Stars in Transition.} ASP Conference Series; Vol. 120; 1997; ed. Antonella Nota and Henny Lamers (1997), p.26
\bibitem[\protect\citeauthoryear{Woosley et al.}{2007}]{woo07} Woosley, S. E., Blinnikov, S., Heger, A. 2007, Nature, 450, 390 
\bibitem[\protect\citeauthoryear{Zampieri et al.}{2005}]{zam05} Zampieri, L., Mucciarelli, P., Pastorello, A., Turatto, M., Cappellaro, E., Benetti, S. 
2005, MNRAS, 364, 1419 
\bibitem[\protect\citeauthoryear{Zickgraf et al.}{1989}]{zick89} Zickgraf, F.-J., Wolf, B., Stahl, O., Humphreys, R. M. 1989, A\&A, 220, 206 
\end{thebibliography}
\end{document}